\documentclass[11pt]{article}
\pdfoutput=1  % togliere il commento se le figure sono pdf, Blasciarlo se sono eps
\usepackage{graphicx}
\graphicspath{{./figures/}}
\usepackage{appendix}
\usepackage{latexsym,amsmath,amsfonts,amssymb,booktabs}
\usepackage[font=small]{caption}
\usepackage{slashed,upgreek,amscd,cancel,color}
\usepackage{adjustbox}
\usepackage[numbers,compress,square]{natbib}
\usepackage{epsfig,latexsym}
\usepackage[pdfencoding=auto]{hyperref}
\usepackage{url}
\numberwithin{equation}{section}% numera le equazioni seconde le sezioni , e.g. 1.15 invece che consecutivamente; anche le appendici, eq. (A.1) etc. Richiede amsmath
\usepackage{doi}
\usepackage{subcaption}
\definecolor{MyBlue}{rgb}{0.15,0.15,0.70}

\hypersetup{
colorlinks=true,
citecolor=MyBlue,
linkcolor=MyBlue,
urlcolor=MyBlue
}

\input epsf
\usepackage{mathtools}
\usepackage{amssymb}
\usepackage{amsmath}
\usepackage{amsfonts}
\usepackage{upgreek}
\usepackage{latexsym}
\usepackage{stfloats}
\usepackage{afterpage}

\newcommand{\be}{\begin{equation}}
\newcommand{\ee}{\end{equation}}
\newcommand{\beq}{\begin{equation}}
\newcommand{\eeq}{\end{equation}}
\newcommand{\bea}{\begin{eqnarray}}
\newcommand{\eea}{\end{eqnarray}}
\newcommand{\bu}{{\bf{u}}}
\newcommand{\bv}{{\bf{v}}}

\newcommand{\bee}{{\bf{e}}}
\newcommand{\obs}{_{\rm O}}
\newcommand{\BH}{_{\rm BH}}
\newcommand{\SBH}{_{\rm SBH}}
\newcommand{\Gal}{_{\rm G}}
\newcommand{\CMB}{_{\rm CMB}}
\newcommand{\CDM}{_{\rm CDM}}
\newcommand{\PBH}{_{\rm PBH}}

\newcommand{\dd}{\text{d}}
\newcommand{\bx}{{\bf{x}}}
\newcommand{\bn}{{\bf{n}}}

\newcommand{\nn}{{\nonumber}}

\usepackage[stable]{footmisc}

\newcommand{\om}{\omega}

\renewcommand\[{\left[}

\newcommand\ees{\end{eqnarray}}
\newcommand\bees{\begin{eqnarray}}

\definecolor{dgreen}{rgb}{0,0.7,0.0}

\begin{document}
\vspace{0.5cm}

\begin{center}
\Large{\textbf{Polarization of a stochastic gravitational wave background through diffusion by massive structures}} \\[1cm]

\large{Giulia Cusin$^a$, Ruth Durrer$^b$, Pedro G. Ferreira$^a$}
\\[0.5cm]

\small{
\textit{$^a$ Astrophysics Department, University of Oxford, DWB, Keble Road, Oxford OX1 3RH, UK}\\
\vspace{.2cm}
\textit{$^b$ D\'epartment de Physique Th\'eorique and Center for Astroparticle Physics,
Universit\'e de Gen\`eve, Quai E. Ansermet 24, CH-1211 Gen\`eve 4, Switzerland}}

\vspace{.2cm}

\vspace{.2cm}

\vspace{0.5cm}
\today

\end{center}

\vspace{2cm}

\begin{abstract}
The geometric optics approximation traditionally used to study the propagation of gravitational waves on a curved background, breaks down in the vicinity of compact and extended astrophysical objects, where wave-like effects like diffusion and generation of polarization occur. 
We provide a framework to study the generation of polarization of a stochastic background of gravitational waves propagating in an inhomogeneous universe. The framework is general and can be applied to both cosmological and astrophysical gravitational wave backgrounds in any frequency range. We derive an order of magnitude estimate of the amount of polarization generated for cosmological and astrophysical backgrounds, in the frequency range covered by present and planned gravitational wave experiments. For an astrophysical background in the PTA and LISA band, the amount of polarization generated is suppressed by a factor $10^{-4}$ ($10^{-5}$) with respect to anisotropies. For a cosmological background we get an additional $10^{-2}$ suppression.  We speculate on using our approach to map the distribution of (unresolvable) structures in the Universe.
\end{abstract}

\newpage

%\newpage
\tableofcontents

\vspace{.5cm}
\newpage

%%%%%%%%%%%%%%%%%%%%%%%%%%%%%%%%%%%%%%%%%%%%%%%%%%
%%%%%%%%%%%%%%%%%%%%%%%%%%%%%%%%%%%%%%%%%%%%%%%%%%
\section{Introduction}
%%%%%%%%%%%%%%%%%%%%%%%%%%%%%%%%%%%%%%%%%%%%%%%%%%
%%%%%%%%%%%%%%%%%%%%%%%%%%%%%%%%%%%%%%%%%%%%%%%%%%

%
Several diffuse stochastic backgrounds of different types of radiation, arising from the  incoherent superposition  of signals from resolved and unresolved sources, have  been observed in astronomy. In particular, backgrounds of electromagnetic radiation include the cosmic microwave background (CMB) with its black body spectrum~\cite{Penzias:1965wn}, cosmic infrared background (CIB) from stellar dust~\cite{2001ARA&A..39..249H} and the extragalactic background light made up of all the electromagnetic radiation emitted by stars, galaxies, galaxy clusters etc. since their formation~\cite{1967ApJ...148..377P,1991Natur.353..315S}. Similarly, there should exist a neutrino background~\cite{2006ARNPS..56..137H} and a background of gravitational waves (GW).  

We can distinguish between a stochastic background of gravitational radiation of cosmological  origin (CGWB) and one of astrophysical origin (AGWB). In the standard cosmological model~\cite{PeterUzan2005}, the existence of a primordial GW background from the amplification of vacuum quantum fluctuations is a generic prediction of any inflationary phase. Gravitational waves may also be produced at the end of inflation during the reheating phase (see e.g. Ref.~\cite{Dufaux:2007pt} for an analytic and numerical study). More speculative sources of a GW background produced at early times include pre big-bang models, cosmic strings~\cite{Vilenkin:1981bx, Hogan:1984is, Vachaspati:1984gt, Caldwell:1991jj, Kuroyanagi:2016ugi}, first order phase transitions in the early universe ~\cite{Caprini:2009fx, Caprini:2015zlo}, magnetic fields \cite{Caprini:2001nb}; see Refs.~\cite{Binetruy:2012ze,Caprini:2018mtu} for a review on those topics and  Refs.~\cite{Maggiore:1999vm, Buonanno:2014aza} for more broader introductions. In addition, an astrophysical background results from the superposition of a large number of resolved and unresolved sources from the onset of stellar activity until today. The nature of the AGWB may differ from its cosmological counterpart, which is expected to be (roughly) stationary, unpolarized, statistically Gaussian and isotropic, by analogy with the cosmic microwave background. Many different astrophysical sources may contribute to the AGWB, including black holes and neutron star mergers~\cite{TheLIGOScientific:2016wyq, Regimbau:2016ike, Mandic:2016lcn, Dvorkin:2016okx, Nakazato:2016nkj, Dvorkin:2016wac, Evangelista:2014oba}, supermassive black holes~\cite{Kelley:2017lek}, neutron stars~\cite{Surace:2015ppq, Talukder:2014eba, Lasky:2013jfa}, stellar core collapse~\cite{Crocker:2017agi, Crocker:2015taa} and population III binaries~\cite{Kowalska:2012ba}.

The recent detection by the Advanced Laser Interferometric Gravitational-wave Observatory (LIGO) of the gravitational wave sources GW150914~\cite{Abbott:2016blz} provided the first observation of the merging of a binary black hole system. Over the last three years, in total, five  detections and one sub-threshold candidate from binary black hole merger events
have  been  reported:    GW150914  \cite{Abbott:2016blz},   GW151226  \cite{Abbott:2016nmj},
LVT151012 \cite{TheLIGOScientific:2016pea}, GW170104 \cite{Abbott:2017vtc}, GW170608 \cite{Abbott:2017gyy} and GW170814  \cite{Abbott:2017oio}. 
 Following these observations, the rate and mass of coalescing binary black holes appear to be greater than many previous expectations. Moreover, the LIGO and Virgo\footnote{{\tt https://www.ego-gw.it/public/about/whatIs.aspx}} collaboration very recently detected  a  new
gravitational-wave  source, GW170817: the  coalescence  of  two  neutron  stars \cite{TheLIGOScientific:2017qsa}.  The merger rate of binary neutron stars
estimated from this event suggests that distant binary neutron stars create a significant contribution to the AGWB which will  add  to  the  background  from  binary  black  holes,  increasing  the  amplitude  of  the  total  astrophysical  background  relative  to  previous  expectations.   In  the  Advanced  LIGO-Virgo  frequency
band most sensitive to stochastic backgrounds (near 25 Hz), the predicted amplitude of the total background is $\Omega_{\rm GW}(f=25 \text{Hz})=1.8_{-1.3}^{+2.7}\cdot 10^{-9}$ with 90\% confidence level compared to $\Omega_{\rm GW}(f=25 \text{Hz})=1.1_{-0.7}^{+1.2}\cdot 10^{-9}$ from binary black holes alone.  Assuming the most probable rate
for compact binary mergers,  in \cite{Abbott:2017xzg} they find that the total background may be detectable with a signal-to-noise-ratio  of  3  after  40  months  of  total  observation  time.      
This improves bounds on the stochastic background obtained from the analysis of big-bang nucleosynthesis~\cite{Maggiore:1999vm, Allen:1996vm},  and of the cosmic microwave background~\cite{Smith:2006nka,Henrot-Versille:2014jua} at 100~Hz. At low frequencies, Pulsar Timing Arrays (see below) give a bound  $\Omega_{\rm GW}<1.3\times10^{-9}$ for $f=2.8 \times 10^{-9}$~Hz~\cite{Shannon:2013wma}. % 
The possibility of measuring and mapping the gravitational wave background is discussed in Refs.~\cite{Allen:1996gp, Cornish:2001hg, Mitra:2007mc, Thrane:2009fp, Romano:2015uma, Romano:2016dpx} while different methods employed by LIGO and LISA (Laser Interferometer Space Antenna) to reconstruct an angular resolved map of the sky are presented in Ref.~\cite{TheLIGOScientific:2016xzw}. An analogous discussion for  Pulsar Timing Arrays  can be found in Refs. \cite{Mingarelli:2013dsa,Taylor:2013esa, Gair:2014rwa}.

The observational landscape is growing and covers large bands of frequencies; see e.g. Ref.~\cite{Moore:2014lga} for a review\footnote{The associated code {\tt http://rhcole.com/apps/GWplotter/} allows one to generate plots of  noise curves for many detectors and associated target sources.}.  \textcolor{black}{At extremely low frequencies $\sim 10^{-16}$ Hz bounds come mainly from the analysis of CMB B-modes} while at  low frequency of order $10^{-10}-10^{-6}$~Hz, there are pulsar timing arrays such as the radio telescope Parks Pulsar Timing Array\footnote{{\tt http://www.atnf.csiro.au/research/pulsar/ppta/}} (PPTA), the Large European Array for Pulsar Timing\footnote{{\tt http://www.leap.eu.org}} (LEPTA) and the future International Pulsar Timing Array\footnote{{\tt http://www.ipta4gw.org}} (IPTA). At low frequencies (typically $10^{-6}-1$~Hz) detection relies on space-borne detectors, such as the Laser Interferometer Space Antenna\footnote{{ www.lisamission.org}} (LISA) and the evolved Laser Interferometer Space Antenna\footnote{{\tt https://www.elisascience.org}} (eLISA) selected by ESA to be launched around 2030. High frequency (typically $1-10^{5}$~Hz) observations rely on ground-based detectors, such as  LIGO and its advanced configuration (aLIGO), Virgo, the Einstein Telescope\footnote{{\tt http://www.et-gw.eu}} (ET) \textcolor{black}{or its American counterpart, the Cosmic Explorer (CE)} \cite{Evans:2016mbw}. This spectrum covers most of the theoretical predictions.

Traditionally the energy density of GW of an astrophysical background has been modelled and parameterized under the assumption that both our universe and the distribution of sources are homogeneous and isotropic, see e.g. Refs.~\cite{Regimbau:2011rp,Dvorkin:2016okx}. These assumptions can be relaxed in order to take into account that astrophysical sources are located in cosmic structures that indeed have a distribution that can be computed in a given cosmological model. Therefore the  energy flux from all astrophysical sources (resolved and unresolved) is not constant across the sky and depends on the direction of observation. In \cite{Cusin:2017fwz}   an analytic framework is presented to describe and compute the anisotropies in the observed energy density of the AGWB, taking into account  the presence of inhomogeneities in the matter distribution and in the geometry of the observed universe.  In \cite{Cusin:2017mjm} an alternative (more geometrical) derivation of the result of \cite{Cusin:2017fwz} is presented, and first numerical predictions for the amplitude of anisotropies for the contribution of the background coming from black hole mergers can be found in \cite{Cusin:2018rsq}. For an astrophysical background, the origin of anisotropies is two fold: first, sources are not isotropically distributed and second, a GW signal, once emitted, is deflected by structures. For a background of cosmological origin, lensing by large scale structures is the main source of anisotropy and actually, CGWB anisotropies are a tracer of CMB temperature anisotropies (see e.g. \cite{Geller:2018mwu} for the case of a CGWB from phase transition). First constraints on the anisotropy have been obtained by PTA~\cite{Taylor2015,Sesana2008},  and from the  first observing run of advanced LIGO~\cite{TheLIGOScientific:2016dpb}. 

In the framework developed in Refs.~\cite{Cusin:2017fwz, Cusin:2017mjm,Cusin:2018rsq} the propagation of GW from the source to the observer is computed in the geometric optics approximation. This is also traditionally done for photons, e.g. to compute CMB temperature anisotropies  or fluctuations of the galaxy distribution see e.g. \cite{Bonvin:2011bg}. The geometric optics approximation is justified as long as the wavelength of gravitons (or photons) is much smaller than the length scale given by the Kretschmann scalar of the metric describing the region of spacetime where the graviton (photon) propagates. In particular, for both gravitons and photons, this approximation is well motivated on cosmological distances, where spacetime is well described by a Friedmann-Lema\^{\i}tre metric with scalar perturbations. In the geometric optics approximation, the polarization tensor is parallel transported along a geodesic: an initially unpolarized background stays unpolarized when it propagates in an inhomogeneous medium. In other words, geometric optics can not describe the generation of polarization. However, in the vicinity of a compact object, wave-like effects are present and sizable and the geometric optics approximation does not capture  interesting effects like diffraction and the generation of polarization. In this regime a wave-like description of GW propagation is necessary. We emphasize that this result holds for GW where the wavelength of gravitons is typically much larger than the one of CMB or infra-red photons. Therefore, when studying the propagation in a highly inhomogeneous medium, the analogy between the behavior of electromagnetic and gravitational radiation may break down.

In this paper we study the generation of polarization of an (initially unpolarized) GW background by diffusion through massive structures. We provide a  framework to describe the generation of polarization and we derive an order of magnitude estimate of the polarization created by the interaction of GWs with matter, in different frequency bands. Our treatment bears several analogies with the creation of polarization by Thomson scattering of the CMB. The role played by electrons in the Thomson scattering is played here by compact and extended astrophysical objects (massive structures). In particular, as for the CMB,  it is the combined effect of anisotropies in the energy density of the background and of the polarization-dependence of the cross section effectively describing the process of scattering, that are responsible for the generation of polarization.  The amplitude of the polarization generated depends on several factors: on the abundance of scattering centers (number density of massive structures), on the relative amplitude of anisotropies of the radiation impinging on a scattering center,  on the wavelength of the GW and on the size of the integrated cross section describing the scattering off a given type of massive structure. This latter effect, in turn, depends on the geometrical properties of the astrophysical target (radius and mass) and on the wavelength. Interestingly, as we will explain in detail, by measuring the  polarization of a given component of the background at different frequencies, it may possible to set constraints on the abundance of some exotic and unresolvable sources in the Universe.

This work is structured as follows. In section \ref{general} we illustrate the general idea underlying our framework. In particular, in sections \ref{wave} and \ref{sec cross} we explain that beyond the geometric optics approximation, the interactions of GW with structures can be described as a diffusion process characterized by an polarization-dependent effective cross section. In section \ref{visibility} we introduce the visibility function for a multi scattering process. In the remaining part of section \ref{general} we introduce the main ingredients needed to fully characterize the generation  of polarization: in section \ref{Stokes} we define Stokes parameters for a GW background, in \ref{angular} we study the angular dependence of a (single) scattering event of a GW off an massive structure and in \ref{after} we compute the Stokes parameters after the scattering. Finally in \ref{polarization tensor} we put all these ingredients together to derive an expression for the polarization tensor. In section \ref{analytical} we provide analytical approximations for the polarization tensor of a GW background (both astrophysical and cosmological) as a sum of contributions of scattering off different types of massive structures. In section \ref{visibility 1} we present order of magnitude estimates of the amount of polarization generated by diffusion for cosmological and astrophysical backgrounds at different frequencies. Finally, in \ref{conclusion} we discuss our results and future perspectives.

\section{General framework}\label{general}

As for the CMB, the generation of polarization of a GW background occurs due to the combined effect of : 
\begin{itemize}
\item the presence of anisotropies in the energy density of the background; 
\item the dependence of the effective scattering cross section of GWs by massive structures on incoming direction and polarization. 
\end{itemize}
In the following we explain in which regime the geometric optics approximation to describe GW propagation breaks down. This happens in the vicinity of a mass distribution in the form of  compact (black hole) or extended objects, we denote them 'massive structures'. We then explain how to treat diffraction effects as an effective scattering process. \textcolor{black}{When studying the angular dependence of the scattering process and computing the Stokes parameters after a scattering, sections \ref{angular} and \ref{after} respectively, we make use of an approach similar to the standard description of CMB polarization in terms of Stokes parameters. In particular, we follow the pedagogical derivations presented in the textbook \cite{Maggiore:2018sht}. }

\subsection{Wave-like effects in GW propagation}\label{wave}

We write the metric describing the geometry of the spacetime as 
\be
g_{\mu\nu}=\bar{g}_{\mu\nu}+h_{\mu\nu}\,,
\ee
where with an overbar we denote the background metric and $h_{\mu\nu}$ is a rapidly  varying small perturbation on the top of it.  Within linear perturbation theory, $h_{\mu\nu}$ satisfies the equation
\be\label{eom}
\bar{\square}h_{\mu\nu}+2 \bar{R}_{\mu\alpha\nu\beta}h^{\alpha\beta}=0\,,
\ee
where $\bar{R}_{\mu\alpha\nu\beta}$ is the Riemann tensor of the background metric and $\bar{\square}$ is the d'Alembertian of the background metric. 

The geometric optics approximation consists in writing 
\be
h_{\mu\nu}(x)=\left(A_{\mu\nu}(x)+\epsilon B_{\mu\nu}(x)+\dots\right)e^{i\theta(x)/\epsilon}\,,
\ee
inserting it in (\ref{eom}) and keeping leading order terms in $\epsilon$. It is easy to verify that, in this approximation, the second term in eq. (\ref{eom}) is systematically discarded. It follows that geometric optics is valid in the regime
\be\label{go}
\frac{1}{\lambda^2}\gg \sqrt{\bar{R}_{\mu\alpha\nu\beta}\bar{R}^{\mu\alpha\nu\beta}}
\equiv \sqrt{K}\,,
\ee
where  {$\lambda \sim (\partial_0\theta)^{-1}$  denotes  a typical wavelength of the GW} and the quantity on the right hand side of this equation is  the square root of the Kretschmann scalar of the metric $\bar{g}_{\mu\nu}$. 

The vicinity of a compact object can be approximately described by a Schwarzschild metric with Kretschmann scalar $K(r)=12r^2_s/r^6$, where $r_s=2GM$ is the Schwarzschild radius of the mass $M$. The condition (\ref{go})  defines a region around the object within which wave-like effects are present, given by $r_s\leq r\leq R_{\lambda}$ where  $R_{\lambda}$ is defined by $K(R_{\lambda}) = \lambda^{-4}$, a length scale which depends on the mass of the object and on the wavelength of the GW under consideration.  For a fixed mass, the size of this region obviously increases with  increasing wavelength. Analogous conclusions hold for the case of an extended object: in this case $R_{\lambda}$ depends on both the mass and the radius of the object through a combination of these two quantities (compactness). In section \ref{computation diffraction}, we will explicitly work out the wave-effect region for different types of astrophysical objects and for the wavelength range  of current and planned GW experiments have access to.

\subsection{Effective treatment of scattering}\label{sec cross}

Let us assume we have a GW  impinging on a black hole (extended object) of mass $M$ (and radius $R$),  with impact parameter $b>r_s$ ($b>R$). The literature on  gravitational scattering of massless waves of various spin is broad and stretches back over forty years; see the monograph \cite{BHmonography} for an extensive treatment of the subject. Over the years, several authors using different methods (see Refs.~\cite{Matzner:1977dn, Westervelt:1971pm, Peters:1976jx, Sanchez:1976fcl, DeLogi:1977dp, Doran:2001ag, Holstein:2006bh,  Dolan:2007ut, Guadagnini:2008ha}) have shown that the differential cross section (summed over polarizations) describing wave-scattering depends on  the spin $s$ of the scattered field as 
\be
\frac{1}{8(MG)^2}\frac{d\sigma}{d\Omega}=
\begin{dcases*}
\frac{1}{\sin^4\theta/2}\,,&\, $s=0$\,,\\
\frac{\cos^2\theta/2}{\sin^4\theta/2}\,,&\, $s=1/2$\,,\\
\frac{\cos^4\theta/2}{\sin^4\theta/2}\,,&\, $s=1$\,,\\
\frac{\cos^8\theta/2+\sin^8\theta/2}{\sin^4\theta/2}\,,&\, $s=2$\,,\label{cross1}
\end{dcases*}
\ee
where $d\Omega\equiv d\cos\theta d\phi$ and $\theta$ is the scattering angle.  The cross section for $s=2$ is the result of at least four separate studies. The first derivation was carried out in \cite{Westervelt:1971pm} applying perturbation theory to the linearized gravitational equations. The author of Ref.~\cite{Peters:1976jx} finds the same result via a Green's function approach while in Ref.~\cite{DeLogi:1977dp} Feynman diagram techniques are employed. Finally the author of Ref.~\cite{Dolan:2007ut}  finds again the same result using partial wave methods and improves on previous work \cite{Matzner:1977dn} which uses the same techniques.\footnote{We emphasize that the gravitational result is somewhat anomalous since it does not follow the same general rule $d\sigma/d\Omega=M^2 \cos^{4s}\theta/2/\sin^4\theta/2$ as other fields. As explained in \cite{Dolan:2007ut}, the origin of the extra term $\sin^4\theta/2$  is a direct consequence of the non-conservation of helicity in gravitational-wave scattering. Helicity is not conserved because axial and polar waves are scattered in a different way. }

We observe that for arbitrary spin, the cross section diverges in the forward direction, i.e. 
\be
\frac{d\sigma}{d\Omega}\propto \frac{1}{\theta^4}\qquad\text{for}\quad \theta\rightarrow 0\,.
\ee
This divergence is due to the long-range nature of gravitational interactions and is present in every scattering process of a \emph{charged} massless wave (or particle) in the Coulomb-like potential  generated by a charged object.\footnote{The Rutherford cross-section can be considered as the electromagnetic counterpart of (\ref{cross1}) and presents the same type of divergence in the forward direction.}
This divergence is physical and it is due to the fact that the Coulomb potential, that is used to model the gravitational potential of a massive object in a galaxy, is long-ranged. Nevertheless  a natural cut-off scale is present in the problem under study. The far-field relation between deflection angle and impact parameter is given by
\be
\theta\approx \frac{2 r_s}{b}\,.
\ee
As we will see in detail in section \ref{computation diffraction}, for both compact and extended objects  wave-like effects occur in a region of space around the object of radius $r\leq R_{\lambda}$ where the parameter $R_{\lambda}$  depends on both the GW wavelength under consideration and the geometrical properties of the object. This sets an upper bound on the impact parameter $b_{\text{max}}=R_{\lambda}$ and correspondingly a lower bound on the deflection angle 
\be
\theta_{\text{min}}=\frac{2 r_s}{b_{\text{max}}}\,.
\ee

The results (\ref{cross1}) are found assuming an unpolarized incoming flux and summing over the final polarization states. The polarization-dependent differential cross section for gravitational wave scattering is given by, see e.g. \cite{Holstein:2006bh, Guadagnini:2008ha} 
\be\label{cross}
\frac{d\sigma^P}{d\Omega}=(MG)^2\frac{1}{\sin^4\theta/2}|e_{ij}'(\bn') e^{*ij}(\bn)|^2\,,
\ee
where $\bn'$ and $\bn$ are the directions of the incoming and outgoing gravitons, respectively and likewise fo the polarizations, $e'_{ij}$ and $e_{ij}$.  The angle $\theta$ is the scattering angle, i.e. $\bn\cdot \bn'=\cos\theta$.  Using the results of appendix \ref{basis}, it is easy to verify that the sum over polarizations of eq. (\ref{cross}) gives back eq. (\ref{cross1}) for $s=2$.

\subsection{Visibility function}\label{visibility}

We recall that if a particle scatters with a cross-section $\sigma$ off an ensemble of targets with number density $n$, its mean free path is $\ell=1/(n\sigma)$. In our case, gravitons scatter off astrophysical objects with physical number density $n_{\text{ph}}$ and the relevant cross-section for scattering off different astrophysical objects is discussed in section \ref{sec cross}. One usually  defines the optical depth due to scattering in the time interval $[\eta_2,\eta_1]$ by
\be\label{tau}
\tau(\eta_1, \eta_2)=\int_{\eta_2}^{\eta_1} d\eta\, n_{\text{ph}} (\eta) \sigma(\eta) a(\eta)\,,
\ee 
where $\eta$ is conformal time and $\eta_1>\eta_2$. The probability that a graviton does not scatter off an astrophysical object in the conformal time interval $[\eta_2, \eta_1]$ is given by 
\be
P(\eta_1, \eta_2)=e^{-\tau(\eta_1, \eta_2)}\,.
\ee
The probability density that a graviton observed at a time $\eta_1$ has undergone a scattering in the interval $[\eta_2, \eta_2+d\eta_2]$ is given by $P(\eta_1, \eta_2+d\eta_2)-P(\eta_1, \eta_2)$. We define the \emph{visibility function} $V(\eta_1, \eta_2)$ as follows:
\be
V(\eta_1, \eta_2) d\eta_2\equiv P(\eta_1+\eta_2+d\eta_2)-P(\eta_1, \eta_2)\,.
\ee
It is easy to verify that
\begin{align}\label{visibilityfun}
V(\eta_1, \eta_2)&=\frac{d}{d\eta_2}e^{-\tau(\eta_1, \eta_2)}=-e^{-\tau(\eta_1, \eta_2)}\frac{d\tau(\eta_1, \eta_2)}{d\eta_2}\,,
\end{align}
with 
\be\label{derivative}
\frac{d\tau(\eta_1, \eta_2)}{d\eta_2}=-n_{\text{ph}} (\eta_2)\sigma a(\eta_2)\,.
\ee
Hence 
\begin{align}
V(\eta_1, \eta_2)=e^{-\tau(\eta_1, \eta_2)}n_{\text{ph}} (\eta_2)\sigma a(\eta_2)\,.
\end{align}
We are mostly interested in $V(\eta)=V(\eta_0, \eta)$ since we observe gravitons today. The quantity 
\be
\int_{\eta}^{\eta_0}d\eta'\, V(\eta_0, \eta')=e^{-\tau(\eta_0, \eta)}\Big|_{\eta}^{\eta_0}=1-e^{-\tau(\eta_0, \eta)}\,,
\ee
by construction is the probability that a graviton observed today has scattered in the time interval $[\eta, \eta_0]$. 

A similar treatment is used for the CMB, \textcolor{black}{see e.g. chapter 20 of \cite{Maggiore:2018sht} which we have followed here},  by substituting $n_{\text{ph}}\rightarrow n_e$ and $\sigma \rightarrow \sigma_{T}$, where $n_e$ is the density of electrons and $\sigma_T$ is the Thomson cross section. In the case of standard recombination, the visibility function is peaked around recombination. Indeed, before recombination, $\tau$ is very large and $V$ is exponentially suppressed. Much later, $d\tau(\eta_0, \eta)/d\eta$ is small because the density of free electrons is small. The width of  the maximum of the visibility function gives the thickness of the large scattering surface.
% (see Gorbunov-Rubakov (2011)).

 For a GW background  the distribution of scattering centers is extended in redshift and thus the situation is different. In that case we expect the visibility function to be much broader and peaked around a redshift at which most of the astrophysical objects are expected to be located. This is similar to what happens during reionization for the CMB. 
Moreover, for the CMB, since $\tau(\eta_0, \eta_0)=0$ and at early times, say $\eta=0$,  $\tau(\eta_0, 0)\rightarrow \infty$, the visibility function satisfies
\be
\int_0^{\eta_0}d\eta\, V_{\CMB}(\eta_0, \eta)= e^{-\tau(\eta_0, \eta)}\Big|_0^{\eta_0}=1\,.
\ee
Thus $V_{\CMB}(\eta)$ is the normalized probability function that a photon observed today has scattered at conformal time $\eta$.  This is as expected since the total probability that a CMB photon scatters before impinging on an observer equals to one. This property is not, however, true for a GW background.

\subsection{Stokes parameters for GW background}\label{Stokes}

In this section we introduce the Stokes parameters to describe the intensity and polarization of a GW background, see also \cite{Romano:2016dpx, Seto, Gubitosi:2016yoq, Kato:2015bye}.  For a single monochromatic plane wave propagating in direction $\bn$, we have that
\be
\tilde{h}_{ij}(f, \bn)=\tilde{h}_{+}(f, \bn) e^+_{ij}(\bn)+\tilde{h}_{\times}(f, \bn) e^{\times}_{ij}(\bn)\,,
\ee
where the expansion coefficients $h_{+, \times}$ are complex-valued functions and $e^{+,\times}_{ij}$ is the $(+, \times)$ polarization basis (see appendix \ref{basis} for detailed definitions). 
We can introduce a polarization tensor as
\be
P_{ijkl}=\tilde{\mathcal{P}}_{ab} e_{ij}^{a}e_{ij}^{b}\,, \quad \text{with} \quad \tilde{\mathcal{P}}_{ab}=\tilde{h}_a^*\tilde{h}_b\,,
\ee
where $a, b=(+, \times)$. The tensor $\tilde{\mathcal{P}}_{ab}$ is a hermitian $2 \times 2$ matrix and therefore can be written as 
\begin{equation}\label{Pdef0}
\tilde{\mathcal{P}}_{a b}(\bn, f)=\frac{1}{2}\left[ I(\bn, f)\,\sigma_{ab}^{(0)}+U(\bn, f)\,\sigma_{ab}^{(1)}+V(\bn, f)\,\sigma_{ab}^{(2)}+Q(\bn, f)\,\sigma_{ab}^{(3)}\right]\,,
%=& \frac{1}{2} \,I(\bn, f) \,\sigma_{ab}^{(0)}+\mathcal{P}_{ab}(\bn, f)\,,
\end{equation}
where $\sigma^{(\alpha)}$ with $\alpha=1\,, 2\,, 3$ denote the Pauli matrices and $\sigma^{(0)}=1_2$ (i.e. the $2\times 2$ identity matrix).  The objects $I\,, U\,, Q\,, V$ are  four real functions of the  GW direction  ${\bf{n}}$ and are the Stokes parameters. In terms of the polarization coefficients of the GW, the Stokes parameters are given by
\begin{align}\label{Stokesd}
I=|\tilde{h}_+|^2+|\tilde{h}_{\times}|^2\,,\qquad Q=|\tilde{h}_+|^2-|\tilde{h}_{\times}|^2\,,\qquad U=2 \text{Re}(\tilde{h}_{+}^*\tilde{h}_{\times})\,,\qquad  V=2 \text{Im} (\tilde{h}_+^* \tilde{h}_{\times})\,.
\end{align}
The Stokes parameter $I$ is simply the intensity of the  GW, $Q$ is the difference between the intensity of radiation polarized along $e_{ij}^{+}$ and $e_{ij}^{\times}$ (and $U$ is the same in a frame rotated\footnote{\label{foot:U}Note that for spin 1 particles, photons, $U$ describes the polarisation rotated by $\pi/4$ but for gravitons a rotation by $\pi/4$ simply exchanges  $h_+$ and $h_{\times}$, hence $Q\mapsto -Q$.} by $\pi/8$). 
The parameter $V$ describes a phase difference between $h_{+}$ and $h_{\times}$ which results in circular polarization.  Using eqs. (\ref{lov}) and (\ref{lov1}), we can rewrite these parameters in terms of the left and right handed polarization basis defined as 
\bees\label{RL}
e^R_{ij}(\bn)&=&\frac{1}{\sqrt{2}}\left(e_{ij}^{+}(\bn)+i\, e_{ij}^{\times}(\bn)\right)\,,\\
e^L_{ij}(\bn)&=&\frac{1}{\sqrt{2}}\left(e_{ij}^{+}(\bn)-i\, e_{ij}^{\times}(\bn)\right)\,. \label{LR}
\ees
We obtain
\begin{align}
I=|\tilde{h}_L|^2+|\tilde{h}_{R}|^2\,,\qquad Q=2 \text{Re}(\tilde{h}_{R}^*\tilde{h}_{L})\,,\qquad U=2 \text{Im}(\tilde{h}_{R}^*\tilde{h}_{L})\,,\qquad  V=|\tilde{h}_R|^2-|\tilde{h}_{L}|^2\,.
\end{align}
It is useful to introduce the following  tensor
\begin{align}\label{Pdef}
\mathcal{P}_{a b}(\bn, f)%&=\frac{1}{2 I(\bn, f)}\left[U(\bn, f)\,\sigma_{ab}^{(1)}+V(\bn, f)\,\sigma_{ab}^{(2)}+Q(\bn, f)\,\sigma_{ab}^{(3)}\right]\\
&=\left[\mathcal{U}(\bn, f)\,\sigma_{ab}^{(1)}+\mathcal{V}(\bn, f)\,\sigma_{ab}^{(2)}+\mathcal{Q}(\bn, f)\,\sigma_{ab}^{(3)}\right]\,,
%=& \frac{1}{2} \,I(\bn, f) \,\sigma_{ab}^{(0)}+\mathcal{P}_{ab}(\bn, f)\,,
\end{align}
in terms of the normalized Stokes parameters $\mathcal{U}=U/(2I)$, $\mathcal{Q}=Q/(2I)$ and $\mathcal{V}=V/(2I)$. 
We can compute the total amplitude of polarization as 
\be\label{PP}
P(\bn, f)\equiv \frac{1}{\sqrt{2}}\left(\mathcal{P}_{ab}(\bn, f) \mathcal{P}^{ba}(\bn, f)\right)^{1/2}=\sqrt{\mathcal{Q}^2+\mathcal{U}^2+\mathcal{V}^2}\,,
\ee

Under a rotation of an angle $\psi$ around  $\bn$, using the transformation properties of (\ref{Rr}) and (\ref{Ll}), we find (omitting the dependence on frequency)
\bees
\tilde{h}_R(\bn; \psi)&=&e^{i 2\psi} \tilde{h}_R(\bn)\,,\\
\tilde{h}_L(\bn; \psi)&=&e^{-i 2\psi} \tilde{h}_L(\bn)\,.
\ees
It follows that under rotation in the plane orthogonal to $\bn$, the Stokes parameters transform as (omitting the dependence on frequency)
\bees
I(\bn; \psi)&=&I(\bn)\,,\\
V(\bn; \psi)&=&V(\bn)\,,\\
Q(\bn; \psi)+i U(\bn; \psi)&=& e^{- i 4\psi} \left(Q(\bn)+i U(\bn)\right)\,,\\
Q(\bn; \psi)-i U(\bn; \psi)&=& e^{ i 4\psi} \left(Q(\bn)-i U(\bn)\right)\,.
\ees
From this, together with the fact that the Stokes parameters are real, we easily conclude that a pure $Q$ polarisation turns into a pure $U$ polarisation under a rotation by $\pm\pi/8$ and vice versa. This proves footnote~\ref{foot:U}. 
Furthermore, $I$ and $V$ transform as scalars on the sphere under rotations while $Q\pm i U$ are spin-4 objects and can be written as linear combinations of spin-4 spherical harmonics. In particular, one can write
\be\left(
\begin{array}{c}
Q(\bn; \psi)\\
U(\bn;  \psi)
\end{array}
\right)=R(4\psi)\left( \begin{array}{c}
Q(\bn)\\
U(\bn)
\end{array}
\right)
\ee
where $R(4\psi)$ is a rotation matrix describing a rotation around the $\bn$ axis.

\subsection{Angular dependence of the scattering process}\label{angular}

If we have a flux of unpolarized radiation coming from a given direction and impinging on a massive object, the dependence of the cross-section on the polarization tensors generates an outgoing polarized radiation. 
As a first step we compute the net polarization generated when a radial flux of gravitons whose intensity has a given angular dependence $I(\theta', \phi')$ scatters off a massive object at the origin of our reference frame. We consider an incoming graviton whose propagation direction is
\be
\bn'=(\sin\theta'\sin\phi', \sin\theta'\cos\phi', \cos\theta')\,,
\ee
i.e. $\bn'$ is the unit radial vector with angles $(\theta', \phi')$. In the plane transverse to $\bn'$ we introduce two orthonormal vectors 
\begin{align}
\bu'&=(\cos\theta'\sin\phi', \cos\theta'\cos\phi', -\sin\theta')\,,\\
\bv'&=(\cos\phi', -\sin\phi', 0)\,.
\end{align}
Using these vectors and eq. (\ref{defer}) we construct the polarization basis $(e^{\times}_{ij}(\bn'),  e^{+}_{ij}(\bn'))$ for the incoming radiation. Let us choose a reference frame such that the direction of propagation of the outgoing radiation $\bn$, is along the $z$ axis, i.e. $\bn=\bee_z$. Then we can chose $\bu=\bee_x$ and $\bv=\bee_y$ for the polarization basis $(e^{\times}_{ij}(\bn),  e^{+}_{ij}(\bn))$ of the outgoing radiation, using eq. (\ref{defer}). 

If the incoming radiation has $\times$ polarization  then the probability that the outgoing radiation has $\times$ polarization is proportional to\footnote{We recall that the basis $(+, \times)$ is real.}
\be
|e_{ij}^{\times}(\bn')e^{ij\times}(\bn)|^2=4\cos^2\theta'\cos^22\phi'\,.
\ee
Analogously, if the incoming radiation has $\times$ polarization then the probability that the outgoing radiation has $+$ polarization is proportional to 
\be
|e_{ij}^{\times}(\bn')e^{ij+}(\bn)|^2=4\cos^2\theta'\sin^22\phi'\,.
\ee
If the incoming radiation has $+$ polarization then the probability that the outgoing radiation has $\times$ and $+$ polarizations  is proportional to, respectively 
\be
|e_{ij}^{+}(\bn')e^{ij\times}(\bn)|^2=\frac{1}{4}(3+\cos2\theta')^2\sin^22\phi'\,,
\ee
\be
|e_{ij}^{+}(\bn')e^{ij+}(\bn)|^2=\frac{1}{4}(3+\cos2\theta')^2\cos^22\phi'\,.
\ee
If the initial radiation is unpolarized, there is an equal probability that the incoming graviton has $\times$ or $+$ polarization . It follows that the probability that a graviton is scattered out with $\times$ polarization  is given by 
\be
\mathcal{E}_{\times}^2\equiv C(\bn, \bn')\left[|e_{ij}^{\times}(\bn')e^{ij\times}(\bn)|^2+|e_{ij}^{+}(\bn')e^{ij\times}(\bn)|^2\right]\,,
\ee
while the probability that the incoming radiation is scattered out with $+$ polarization is given by
\be
\mathcal{E}_{+}^2\equiv C(\bn, \bn')\left[|e_{ij}^{\times}(\bn')e^{ij+}(\bn)|^2+|e_{ij}^{+}(\bn')e^{ij+}(\bn)|^2\right]\,,
\ee
where $C(\bn, \bn')$ is the prefactor multiplying the polarization-dependent part of the cross-section, eq. (\ref{cross}), and is given by  
\be
C(\bn, \bn')=C \sin^{-4}\frac{\theta'}{2}\,,
\ee
where $C$  is a constant which does not depend on angles and which disappears in the final computation of the polarization tensor. Explicitly, one finds 
\begin{align}
\mathcal{E}_{\times}^2&=C \frac{1}{\sin^{4}\theta'/2} \left[4\left(\cos^8\frac{\theta'}{2}+\sin^8\frac{\theta'}{2}\right)-\frac{1}{2}\cos4\phi'\sin^4\theta'\right]\,,\\
\mathcal{E}_{+}^2&=C \frac{1}{\sin^{4}\theta'/2} \left[4\left(\cos^8\frac{\theta'}{2}+\sin^8\frac{\theta'}{2}\right)+\frac{1}{2}\cos4\phi'\sin^4\theta'\right]\,.\\
\end{align}

\subsection{Stokes parameters after scattering}\label{after}

The Stokes parameter $I$ of the radiation scattered in the $\bee_z$ direction is obtained by integrating $\mathcal{E}_{\times}^2+\mathcal{E}_{+}^2$ over all directions of the incoming radiation, weighted with the intensity of the incoming radiation $I(\theta', \phi')$:
\be
I=C\int d\Omega'\,I(\theta', \phi')\frac{8}{\sin^4\theta'/2}\left(\cos^8\frac{\theta'}{2}+\sin^8\frac{\theta'}{2}\right)\,.
\ee
The Stokes parameter $Q$ is obtained by integrating $\mathcal{E}_{+}^2-\mathcal{E}_{\times}^2$ over directions, again with $I(\theta', \phi')$ as a weight:
\be
Q=C\int d\Omega'\,I(\theta', \phi')\frac{1}{\sin^4\theta'/2}\cos4\phi'\sin^4\theta'\,.
\ee
As explained in section \ref{Stokes}, the Stokes parameter $U$ can be obtained from $Q$ with a rotation around the $\bn$ axis of $\pi/8$\footnote{More precisely, $U$ can be obtained from $Q$ by  projecting on a polarization basis built from $\bu$ and $\bv$ vectors rotated of $\pi/8$ in the plane perpendicular to $\bn$.}
\be
U=-C\int d\Omega'\,I(\theta', \phi')\frac{1}{\sin^4\theta'/2}\sin4\phi'\sin^4\theta'\,.
\ee
The Stokes parameter $V$ is defined as the difference between the intensity polarized $R$ and $L$, see section \ref{Stokes}. We can build a basis $(e_{ij}^L(\bn'), e_{ij}^R(\bn'))$ for the incoming radiation and a basis $(e_{ij}^L(\bn), e_{ij}^R(\bn))$ for the outgoing radiation, starting from the basis $(+, \times)$ and using equation (\ref{RL}). Explicitly 
\bees
e^R_{ij}(\bn)&=&\frac{1}{\sqrt{2}}\left(e_{ij}^{+}(\bn)+i\, e_{ij}^{\times}(\bn)\right)\,,\\
e^L_{ij}(\bn)&=&\frac{1}{\sqrt{2}}\left(e_{ij}^{+}(\bn)-i\, e_{ij}^{\times}(\bn)\right)\,,
\ees
and analogously for the incoming radiation by replacing $\bn\rightarrow \bn'$. 
%\bees
%e^R_{ij}(\bn')&=&\frac{1}{\sqrt{2}}\left(e_{ij}^{+}(\bn')+i\, e_{ij}^{\times}(\bn')\right)\,,\\
%e^L_{ij}(\bn')&=&\frac{1}{\sqrt{2}}\left(e_{ij}^{+}(\bn')-i\, e_{ij}^{\times}(\bn')\right)\,,
%\ees
The probability that an initially unpolarized radiation is polarized $R$ ($L$) after the scattering is given by 
\be
\mathcal{E}_R^2\equiv C\left[|e_{ij}^{R}(\bn')e^{ijR}(\bn)|^2+|e_{ij}^{L}(\bn')e^{ijR}(\bn)|^2\right]\,,
\ee
\be
\mathcal{E}_L^2\equiv C\left[|e_{ij}^{R}(\bn')e^{ijL}(\bn)|^2+|e_{ij}^{L}(\bn')e^{ijL}(\bn)|^2\right]\,,
\ee
respectively. It is easy to verify that
\be
\mathcal{E}_R^2=\mathcal{E}_L^2\,.
\ee
It follows that
\be
V=C\int d\Omega'\,I(\theta', \phi')\frac{1}{\sin^4\theta'/2}\left(\mathcal{E}_R^2-\mathcal{E}_L^2\right)=0\,,
\ee
the scattering of GW radiation off a massive object does not generate circular polarization, in full analogy with the Thomson scattering for electromagnetic radiation. 

Summarizing, we found the following Stokes parameters for the GW radiation along $\bn=\bee_z$, after the scattering of unpolarized radiation off a massive object \footnote{Note that the integral over the scattering angle $\theta'$ has a lower bound $\theta_{\text{min}}$ proportional to the mass of the scatterer and to the frequency of the GW scattering off it, see sections \ref{sec cross} and \ref{integrated}.}
\begin{align}
I&=C\int d\Omega'\,I(\theta', \phi')\frac{8}{\sin^4\theta'/2}\left[\cos^8\frac{\theta'}{2}+\sin^8\frac{\theta'}{2}\right]\,,\label{II}\\
Q&=C\int d\Omega'\,I(\theta', \phi')\frac{1}{\sin^4\theta'/2}\left[\cos4\phi'\sin^4\theta'\right]\,,\label{QQ}\\
U&=C\int d\Omega'\,I(\theta', \phi')\frac{1}{\sin^4\theta'/2}\left[-\sin4\phi'\sin^4\theta'\right]\,,\label{UU}\\
V&=0\,.\label{VV}
\end{align}
The angular factors in square parenthesis in eqs. (\ref{QQ}) and (\ref{UU}) can be expanded in a basis of spherical harmonics with $m=\pm 4$ and $\ell\geq 4$. It follows that an  {\em isotropic} incoming flux of radiation does not generate any net polarization.

Up to now we have chosen the coordinate system such that the direction of propagation of the outgoing radiation was along the $z$-axis. We now rewrite eqs. (\ref{II})-(\ref{VV}) in a coordinate independent (rotationally invariant) way. First  we make use of standard trigonometric identities to rewrite  (\ref{II})-(\ref{UU}) as a functions of $\cos\theta', \sin\theta', \cos\phi', \sin\phi'$ only. Then we introduce two orthonormal vectors in the plane perpendicular to $\bn$, $\bu(\bn)$ and $\bv(\bn)$. For the choice $\bn=\bee_z$ they reduce to $\bu(\bee_z)=\bee_x$ and $\bv(\bee_z)=\bee_y$. With these definitions, we have
\begin{align}
\cos\theta'&=\bn\cdot \bn'\,,\\
\sin\phi'&=\frac{\bn'\cdot \bu(\bn)}{\sqrt{1-(\bn\cdot \bn')^2}}\,,\\
\cos\phi'&=\frac{\bn'\cdot \bv(\bn)}{\sqrt{1-(\bn\cdot \bn')^2}}\,.
\end{align}
After standard simplifications, we find
\begin{align}
I(\bn)&=C\int d^2\bn'\,I(\bn')\frac{2}{\left[1-\bn\cdot \bn'\right]^2}\left[(1+\bn\cdot\bn')^4+(1-\bn\cdot \bn')^4\right]\,,\label{IIn}\\
Q(\bn)&=C\int d^2\bn'\,I(\bn')\frac{4}{\left[1-\bn\cdot \bn'\right]^2}\left[(\bn'\cdot \bv_{\bn})^4+(\bn'\cdot \bu_{\bn})^4-6(\bn'\cdot \bv_{\bn})^2(\bn'\cdot \bu_{\bn})^2\right]\,,\label{QQn}\\
U(\bn)&=C\int d^2\bn'\,I(\bn')\frac{16}{\left[1-\bn\cdot \bn'\right]^2}\left[(\bn'\cdot \bu_{\bn})^3(\bn'\cdot \bv_{\bn})-(\bn'\cdot \bv_{\bn})^3 (\bn'\cdot \bu_{\bn})\right]\,,\label{UUn}\\
V(\bn)&=0\,,\label{VVn}
\end{align}
where we have simplified notation to $\bu_{\bn}=\bu(\bn)$ and $\bv_{\bn}=\bv(\bn)$.  Eqs. (\ref{IIn})-(\ref{VVn}) determine the Stokes parameter of the GW radiation scattered by a black hole into direction $\bn$. 

\begin{figure}
\centering
\includegraphics[scale=0.46]{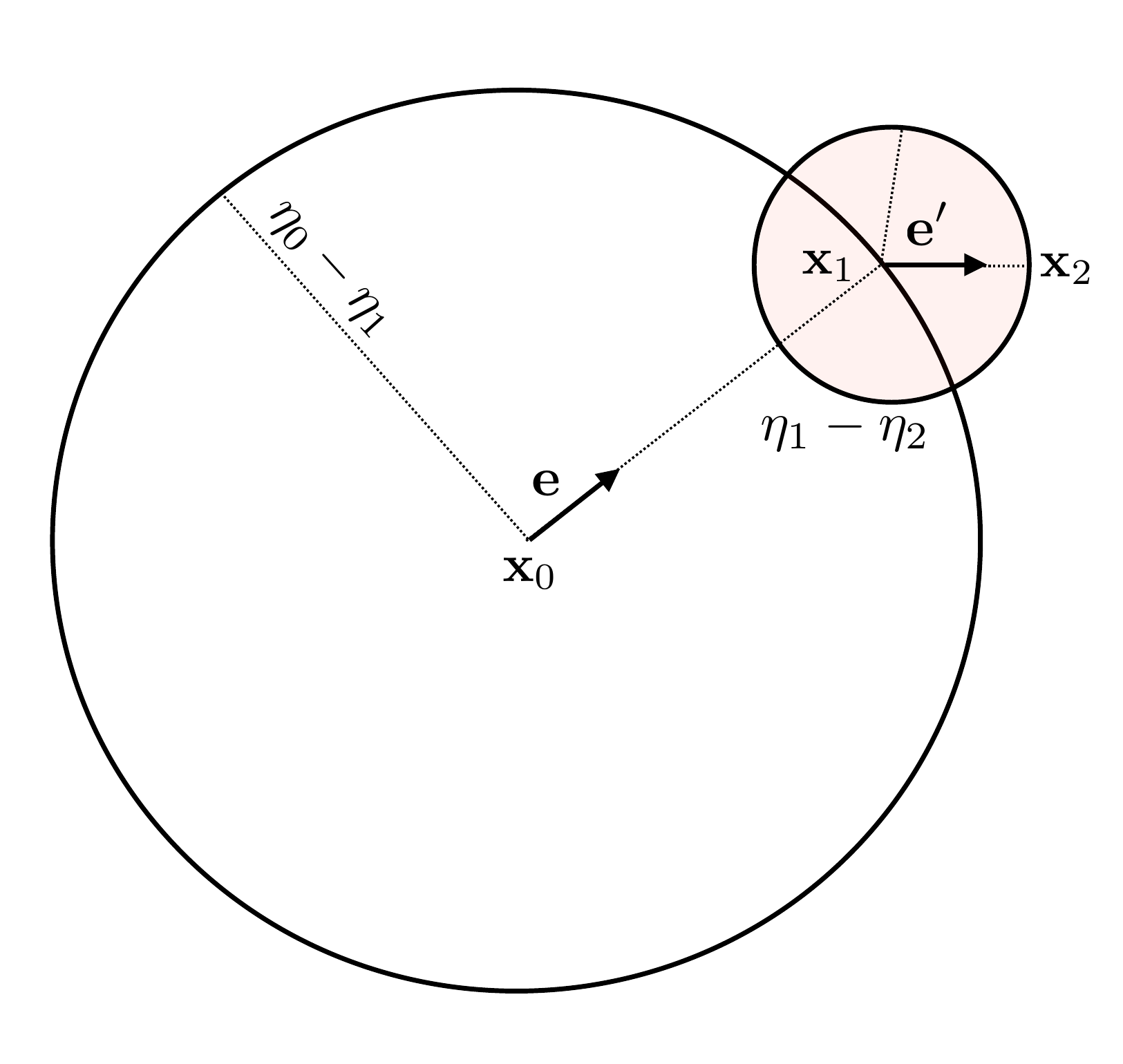}
 \caption{Schematic representation of the scattering processes under study. \textcolor{black}{Readapted from \cite{Maggiore:2018sht}}. }\label{figure}
  \end{figure}

Consider the scattering geometry in Figure \ref{figure} where we neglect lensing effects and the observer receives the outgoing radiation in the direction $\bee=-\bn$. The Stokes parameter which can be measured in the direction of observation $\bee$ are therefore given by
\begin{align}
I(\bee)&=C\int d^2\bee'\,I(\bee')\frac{2}{\left[1-\bee\cdot \bee'\right]^2}\left[(1+\bee\cdot\bee')^4+(1-\bee\cdot \bee')^4\right]\,,\label{IIe}\\
Q(\bee)&=C\int d^2\bee'\,I(\bee')\frac{4}{\left[1-\bee\cdot \bee'\right]^2}\left[(\bee'\cdot \bv_{\bee})^4+(\bee'\cdot \bu_{\bee})^4-6(\bee'\cdot \bv_{\bee})^2(\bee'\cdot \bu_{\bee})^2\right]\,,\label{QQe}\\
U(\bee)&=C\int d^2\bee'\,I(\bee')\frac{16}{\left[1-\bee\cdot \bee'\right]^2}\left[(\bee'\cdot \bu_{\bee})^3(\bee'\cdot \bv_{\bee})-(\bee'\cdot \bv_{\bee})^3 (\bee'\cdot \bu_{\bee})\right]\,,\label{UUe}\\
V(\bee)&=0\,,\label{VVe}
\end{align}
where $\bee'=-\bn'$ and $\bv_{\bee}=\bv(\bee)=\bv(-\bn)$ and $\bu_{\bee}=\bu(\bee)=\bu(-\bn)$ are orthonormal vectors in the plane perpendicular to $\bee$ such that $\bv(\bee=\bee_z)=-\bee_y$ and $\bu(\bee=\bee_z)=-\bee_x$. 

The Stokes parameters still depend on the choice of the directions $\bv_{\bee}$ and $\bu_{\bee}$. This can be avoided by expanding the polarisation into a $E$-mode (gradient type) and $B$-mode   (curl type) components analogous to the total angular momentum decomposition for the CMB (see e.g~\cite{ruthBook}), but we refrain from this further formal development here. 

\subsection{Polarization tensor}\label{polarization tensor}

We split the intensity into an homogeneous and isotropic contribution and an inhomogeneous and anisotropic  one. For gravitational radiation of (observed) frequency $f$ impinging on a scattering center at $\bx$ in the direction $\bee$ and at time $\eta$, we write
\be\label{split}
I(\eta, \bx, \bee, f)=\bar{I}(\eta, f)+\delta I (\eta, \bx, \bee, f)\,. 
\ee
At linear order in $\delta I$, the polarization tensor (\ref{Pdef}) is given by
\be\label{semifinal}
\mathcal{P}_{ab}(\eta_0, \bx_0, \bee, f)=\frac{\int_{0}^{\eta_0}\, d\eta_1\, V(\eta_0, \eta_1)\int d^2\bee'\, \delta I(\eta_1, \bx_1, \bee', f_1)\mathcal{S}_{ab}(\bee, \bee')}{2\bar{I}(\eta_0, \bx_0, f)}\,,
\ee
where $\bx_1=\bx_0+(\eta_0-\eta_1)\bee$  is the position of the scattering center, see figure \ref{figure}, $f_1=(1+z(\eta_1))f$ and we have defined the following quantity which depends only on angles 
\begin{align}\label{SS}
S_{ab}(\bee, \bee')&=\frac{4}{\left[1-\bee\cdot \bee'\right]^2}\left\{
%\sigma^{(0)}_{ab}+
4\left[(\bee'\cdot \bu_{\bee})^3(\bee'\cdot \bv_{\bee})-(\bee'\cdot \bv_{\bee})^3 (\bee'\cdot \bu_{\bee})\right]\sigma_{ab}^{(1)}\right.\nn\\
&\hspace{4 em}\left.\hspace{3 em}+\left[(\bee'\cdot \bv_{\bee})^4+(\bee'\cdot \bu_{\bee})^4-6(\bee'\cdot \bv_{\bee})^2(\bee'\cdot \bu_{\bee})^2\right]\sigma_{ab}^{(3)}\right\}\,. 
\end{align}
In eq. (\ref{semifinal}), $I(\eta_1, \bx_1, \bee', f_1)$ is the intensity of the radiation incident on the scattering center in $(\eta_1, \bx_1)$, from the direction $\bee'$ and at frequency $f_1$.  $\bar{I}(\eta_0, \bx_0, f)$ is the intensity at the observer (averaged over directions). As we will explain in the next section, while polarization is generated only by wave-like effects, the intensity does not vanish in the geometric optics approximation either (i.e. even if diffraction is discarded). 

\section{Analytic expressions for polarization}\label{analytical}

We want to derive an expression for the polarization tensor $\mathcal{P}_{ab}$, eq. (\ref{semifinal}), for a primordial (cosmological) background and for an astrophysical background, both in a cosmological setting (i.e. a Friedmann universe with structures). The ingredients needed are:
\begin{enumerate}
\item the visibility function for different types of scattering, introduced in section \ref{visibility};
\item the intensity of the incoming radiation. 
\end{enumerate}
In section \ref{relation}  we write the intensity of the incoming radiation in terms of the energy density of the background. The energy density at a point $(\eta_1, \bx_1)$ of the space time, and seen in a direction $\bee_1$ by a comoving observer in this position, can be computed by using the Boltzmann approach detailed in appendix \ref{Bol}.

Our approach is as follows. The intensity of the background is computed in the geometric optics approximation, neglecting diffusion effects.  In the Boltzmann equation approach, this corresponds to neglecting the collision  term and solving the Liouville equation with an emissivity part only. We then use this result to compute the polarization, generated by diffusion. Polarization is a purely beyond-geometric optics effect. On the other hand wave-like effects represent a second order correction to the intensity and we neglect them in the present treatment.\footnote{Writing a set of Boltzmann equations for intensity \emph{and} polarization, this approximation would correspond to neglecting the collision term in the equation for the intensity, to solve this independently and use the result as a source to the equation for polarization. This approach is consistent since, as we will see in section \ref{visibility 1}, the polarization generated is very small so that we can neglect the back reaction of polarization on intensity.}

 We consider the standard cosmological framework in which the universe is modeled by a Friedmann-Lema\^{\i}tre-Robertson-Walker (FLRW) universe with Euclidean spatial sections and with scalar perturbations.  In Newtonian gauge, the metric $g_{\mu\nu}$ is given by 
\be\label{FL}
\dd s^2=a(\eta)^2\left[-(1+2\psi)\dd\eta^2+(1-2\phi)\delta_{ij}\dd x^i\dd x^j\right]\,,
\ee
where the metric of the constant time hypersurfaces is
\be
\delta_{ij}\dd x^i\dd x^j=\dd \chi^2+\chi^2\left(\dd\theta^2+\sin^2\theta\dd\phi^2\right)\,,
\ee
in terms of the comoving radial distance $\chi$. The two Bardeen potentials are decomposed as
\begin{equation}
 \psi = \Psi+\Pi,\qquad
 \phi = \Psi-\Pi.
\end{equation}
In the standard $\Lambda$CDM model, the matter content at late time is dominated by cold dark matter (CDM), described by a pressureless fluid, and by the cosmological constant. It follows that the Bardeen potentials,  $\phi$ and $\psi$, are equal, so that $\psi=\phi=\Psi$ and $\Pi=0$. We assume that the galaxies  are all comoving with the cosmic flow.\footnote{The velocity of galaxies is not biased $v(z\,,\bee)=v_{CDM}(z\,,\bee)$.}  To first order in perturbations, the four velocity of the cosmic fluid is given by
\be\label{peculiar}
u^{\mu}\equiv \frac{1}{a}(1-\psi\,, v^i)\equiv \bar u^\mu + \delta u^\mu\,,
\ee
where $v^i$ is the peculiar velocity field. From the matter conservation equation, the galaxy peculiar velocity can be related to the gravitational potential through the Euler equation.

\subsection{The relation between intensity and the background energy density}\label{relation}

As discussed in  \cite{Cusin:2017mjm},  if we want to describe an \emph{inhomogeneous} background, a useful quantity  is the energy density of the background in a given direction, which is quadratic in the signal, does not depend on phases and has a non-vanishing correlation function: 
\be\label{Omega}
\Omega_{GW}(\bee)\equiv \frac{1}{\rho_c}\frac{d^2\rho_{GW}}{d^2\bee}(\bee)\,,
\ee
where $\rho_c$ is the critical density of the universe and $d^2\rho_{GW}/d^2\bee$ is the energy density of the background in the solid angle $d^2\bee$ around $\bee$.
%\footnote{In this context of a GW background, on the right hand side of (\ref{Omega}), the signal has to be interpreted as the superposition of all the signal received at the detector (from different sources) in the direction $\bee$. }
It is useful to introduce the dimensionless energy density per unit of logarithmic frequency, as
\be\label{Omegaf}
\Omega_{GW}(\bee)\equiv \int_0^{+\infty}d\log f\,\, \Omega_{GW}(\bee, f)\,.
\ee
It follows that 
\be
\Omega_{GW}(\bee, f)=\frac{f}{\rho_c}\frac{d^3\rho_{GW}}{d^2\bee df}(\bee, f)\,,
\ee
where $d^3\rho_{GW}$ is the energy density of the background in the solid angle $d^2\bee$ around $\bee$ and in the frequency bin around $f$. Using the standard expression for the energy density in terms of the wave amplitude, see eq. \cite{Maggiore:1900zz},  and recalling that the definition of energy requires an average over several periods of the wave,  we find 
\be\label{preintensity}
\Omega_{GW}(\bee, f)=\frac{c^2}{4G\rho_c}\frac{1}{T_{\obs}}f^3 \sum_A |\tilde{h}_A(f, \bee)|^2\,,
\ee
where the quantity $T_{\obs}$ comes from the time average and represents the period of observation of the detector. The derivation of this result can be found in  \cite{Cusin:2017mjm}. Using the definition of intensity, eq. (\ref{Stokesd}), we find
\be\label{intensityo}
\Omega_{GW}(\bee, f)=\frac{c^2}{4G\rho_c}\frac{1}{T_{\obs}}f^3 I(\bee, f)\,.
\ee

\subsection{Radiation incident on a target}

We use the Boltzmann approach described in appendix  \ref{Bol} to compute the energy density at a point $(\eta_1, \bx_1)$ of the spacetime, from a direction $\bee'$ (see figure \ref{figure} for a representation of the situation under study). We neglect collisions. For a cosmological background we find
\be\label{cosmo fin 1}
\bar{\Omega}_{GW}[\eta_1, \bx_1,  f_1]=\frac{a(\eta_i)^4}{a(\eta_1)^4}\bar{\Omega}_{GW}[\eta_i, f_i]\,, 
\ee
%\rut{[NOTE: I have changed $f_1$ to $f_i$ on the rhs. above and below]}
and 
\begin{align}\label{cosmo fin 2}
&\frac{\delta\Omega_{GW}}{\bar{\Omega}_{GW}}[\eta_1, \bx_1,  f_1, \bee']=\frac{\delta\Omega_{GW}}{\bar{\Omega}_{GW}}[\eta_i, \bx_2, f_i] -4\left\{\phi[\eta_1, \bx_1]-\phi[\eta_i, \bx_2]-\bee'\cdot \bv[\eta_1, \bx_1]+\bee'\cdot \bv[\eta_i, \bx_2]\right\}\,, 
\end{align}
where, to shorten the notation we have defined $\bx_2\equiv \bx_1+\bee'(\eta_1-\eta')$.  For an astrophysical background, keeping the leading term in the perturbation part in eq. (\ref{YY})
\begin{align}\label{astro fin}
\Omega_{GW}[\eta_1, \bx_1, \bee',  f_1]=\frac{f_1}{4\pi\rho_c}\int_0^{\eta_1}d\eta'\, a^4(\eta')\bar{n}_{\Gal}(\eta')\left[1+\delta_{\Gal}[\eta', \bx_2]\right]\int d\vartheta_{\Gal}\, \mathcal{L}_{\Gal}\left[\frac{1+z(\eta', \bx_2)}{1+z(\eta_1, \bx_1)}\,f_1, \vartheta_{\Gal}\right]\,.
\end{align}
This results coincides with the one obtained in  \cite{Cusin:2017fwz, Cusin:2017mjm}. 
In eq. (\ref{astro fin}), $\bar{n}_{\Gal}$ denotes the average density of galaxies, the galaxy overdensity is $\delta_{\Gal}\equiv b \delta_\text{m}$ where $b$ is the bias function, $\delta_\text{m}$ is the matter overdensity  and $\mathcal{L}_{\Gal}$ is the effective luminosity of GW of a galaxy, per units of emitted frequency $f_{\Gal}=(1+z(\eta', \bx_2))/(1+z(\eta_1, \bx_1))\,f_1$ and $\vartheta_{\Gal}$ are a set of parameters effectively describing a given galaxy (mass, metallicity, etc.).  The intensity corresponding to eqs. (\ref{cosmo fin 1}), (\ref{cosmo fin 2}) and (\ref{astro fin}) can be computed using eq. (\ref{intensityo}).

\subsection{Final result for the polarization tensor}\label{end}

We now rewrite the formal result for the polarization tensor, eq. (\ref{semifinal}),  in terms of the building blocks already computed. The visibility function is defined in eq. (\ref{visibilityfun}) and for small optical depth $\tau\ll 1$ it can be approximated as 
\be
V(\eta_0, \eta_1)\simeq-\frac{\partial}{\partial \eta_1}\tau(\eta_1)\equiv -\dot{\tau}(\eta_1)\,,
\ee
where the derivative of the optical depth is given by eq. (\ref{derivative}). In eq. (\ref{semifinal}) for the polarization tensor we found the intensity in terms of the background energy density by using eq. (\ref{intensityo}) and we can use it to give us
\be
\bar{I}[\eta_0, \bx_0, \bee, f]=\frac{4 G \rho_c}{c^2}T_{\obs}\, f^{-3}\bar{\Omega}_{GW}[\eta_0, \bx_0, \bee, f]\,, 
\ee
%\be
%\delta I[\eta_1, \bx_1, \bee', f_1]=\frac{4 G \rho_c}{c^2} T_{\obs}f_1^{-3}\, \delta\Omega_{GW}[\eta_1, \bx_1, \bee', f_1]\,,
%\ee
and 
\be\label{ch}
\delta I[\eta_1, \bx_1, \bee', f_1]=\frac{4 G \rho_c}{c^2} T_{\obs}\bar{f}_1^{-3} \bar{\Omega}_{GW}[\eta_1, \bx_1, f_1]\left[ \frac{\delta\Omega_{GW}}{ \bar{\Omega}_{GW}}[\eta_1, \bx_1, \bee', f_1]-3\frac{\delta f_1}{\bar{f}_1}[\eta_1, \bx_1, \bee']\right]\,,
\ee
where we use that $f_1=(1+z(\eta_1, \bx_1, \bee'))f$ and split the frequency into a background and a perturbation part $f_1=\bar{f}_1+\delta f_1$.\footnote{Perturbations of frequency are proportional to perturbations of redshift $\delta f_1/\bar{f}_1=\delta z/(1+\bar{z})$. }  %analogously for $\bar{I}[\eta_0, \bx_0, \bee, f]$. 
\textcolor{black}{In a cosmological framework, the perturbation of frequency gives a sub-leading contribution to eq. (\ref{ch}) and for simplicity we neglect it in the following.} At first order in perturbations the polarization tensor describing polarization generated by scattering off a given type $(i)$ of targets is then given by 
%\footnote{\textcolor{red}{Here I am considering $C=1$ in (\ref{semifinal}). That's not correct...}}
\be
\mathcal{P}^{(i)}_{ab}(\eta_0, \bx_0, \bee, f)=-\frac{\int_0^{\eta_0}d\eta_1\, \dot{\tau}^{(i)}(\eta_1)\int d^2\bee'\, \delta\Omega_{GW}[\eta_1, \bx_1, \bee', f_1]\mathcal{S}_{ab}(\bee, \bee')}{2 \bar{\Omega}_{GW}[\eta_0, \bx_0, \bee, f]}\,,
\ee
with 
%\rut{[Note: I removed a factor $a^3(\eta_1)$]}
\be\label{final}
\dot{\tau}^{(i)}(\eta_1)=-a(\eta_1)n_s^{(i)}(\eta_1)\sigma_s^{(i)}\,, 
\ee
%\rut{[Note: I removed a factor $a^{-3}(\eta_1)$. Below I again agree with you.]}\\
where $n_s$ is the physical number density of scattering centers of type $i$ (i.e. stellar mass black holes, super massive black hole ...) and $\sigma_s^{(i)}$ is the corresponding integrated cross section. Note that we are assuming that the objects causing diffraction have an isotropic distribution. Taking into account the  angular dependence of the density of targets, $n_s^{(i)}(\eta_1, \bee)$,  is a straightforward generalization of our framework; one just needs to replace the optical depth  in eq. (\ref{final})  with a direction dependent one $\tau(\eta_1)\rightarrow \tau(\eta_1, \bee)$. 

Considering only incoherent scattering, the total polarization is just the sum of the polarization generated by all the different scattering events (see section \ref{coherent}).\footnote{In section \ref{coherent} we will show that the typical separation between scatterers (massive structures) is such that for the frequency range of present and planned GW experiments,  multi-scattering can be treated as  incoherent.} The total polarization tensor is therefore given by 
\begin{align}
\mathcal{P}^{\text{tot}}_{ab}(\eta_0, \bx_0, \bee, f)&=\sum_i \mathcal{P}^{(i)}_{ab}(\eta_0, \bx_0, \bee, f)\nn\\
&=\frac{\int_0^{\eta_0}d\eta_1\,\sum_i \left(n_s^{(i)}(\eta_1, \bee)\sigma_s^{(i)}\right)a(\eta_1)\int d^2\bee'\, \delta\Omega_{GW}[\eta_1, \bx_1, \bee', f_1]\mathcal{S}_{ab}(\bee, \bee')}{2 \bar{\Omega}_{GW}[\eta_0, \bx_0, \bee, f]}\,.
\end{align}
This polarization tensor is a function of the energy density of the GW background which  is given by eqs. (\ref{cosmo fin 1}), (\ref{cosmo fin 2}) and (\ref{astro fin}) in terms of matter and metric perturbations. Polarization is therefore a stochastic quantity which can be characterized statistically in terms of its two-point function, like the  energy density of the GW background in \cite{Cusin:2017fwz}. Moreover, it will cross-correlate with GW energy density and with other cosmological probes such as the galaxy distribution and weak lensing. We will investigate these aspects in  future work.

We have assumed that the flux of the radiation emitted by a given object is unpolarized. For a cosmological background  this is a very good assumption since we expect that $+$ and $\times$ polarization are produced with equal probability.  An astrophysical background is given by the incoherent superposition of signals with random distribution of polarization. The average polarization of the background produced at a given redshift $z$ is therefore vanishing and the variance is proportional to the inverse of the number of sources at that redshift. Moreover, as we will see in section \ref{visibility 1}, the polarization generated by scattering is much smaller than the intensity.

If an initial polarization is sufficiently small, we can just sum it to the polarization generated by interaction with matter. On the other hand, if the initial polarization is sizable, one can no longer neglect the back reaction of polarization on intensity. In this case, in our framework, one needs to go one step further and compute the intensity a background acquires after scattering and compute the new Stokes parameter considering the total intensity in expressions eqs. (\ref{IIe}-\ref{VVe}). The procedure can be reiterated. A more rigorous approach consists in solving a set of Boltzmann equations for intensity and polarization with a collision term which can be derived considering the angular dependence of the cross section (\ref{cross}). This can be done by determining a scattering matrix for all Stokes parameters exactly as for the CMB, see e.g.~\cite{ruthBook}. In the present approach, however, we assume polarization to be very small and neglect its backreaction on intensity.

\section{An estimate of the polarized GW background}\label{visibility 1}

We now want to derive an order of magnitude estimate of the polarization parameter $P$, introduced in eq. (\ref{PP}) that a GW background acquires due to  interactions with matter. We assume that  scattering off massive structures is the only source of polarization. If the GW flux is initially unpolarized, the amount of polarization (averaged over directions) measured by the observer is proportional to the probability of scattering and to the amount of anisotropies of the radiation incident on a  scattering center. Using the results of the previous sections, we can derive an an estimate of the amount of polarization a flux of gravitons produced at $z$ and received today as 
\be\label{QQQ}
P(z)=\frac{\delta \Omega_{GW}}{\bar{\Omega}_{GW}}\times \left[1-e^{-\tau(z)}\right]\,,
\ee
where the optical depth $\tau$ is defined in eq. (\ref{tau}). For comparison, for the CMB, assuming that all the polarization is generated by reionisation and $\tau_{\text{rei}}=0.08$, one finds 
\be\label{QQCMB}
P^{\text{rei}}_{\CMB}=\frac{\delta T}{\bar{T}}\times \left[1-e^{-\tau(z_{\text{rei}})}\right]\simeq 10^{-6}\,.
\ee
%\rut{[NOTE: I don't understand the number with $\delta T/\bar T \sim 10^{-5}$ I get $8.3\times 10^{-7} \simeq 10^{-6}$ ]}

Two main ingredients enter eq. (\ref{QQQ}): the fractional anisotropy of a GW background and the optical depth. This last ingredient depends on the density of scattering centers and on the effective cross section of the scattering process. In this section we derive an order of magnitude estimate for the optical depth for different types of astrophysical objects acting as scattering centers. The optical depth will be a function of the GW frequency $1/\lambda$. For future reference, in table~\ref{bella} we list the wavelength range of (some) current and planned GW experiments. 
\begin{table}
\begin{center}
\begin{tabular}{ccc}
Experiment          &    $\lambda_{\obs}$[pc]  &$\lambda_M$[pc]\\
\hline\\
\vspace{-2 em}\\
LIGO & $10^{-11}-10^{-9}$ & $10^{-10}$ \\
LISA          &  $10^{-9}-10^{-5}$ &$10^{-7}$ \\
PTA        & $10^{-5}-10^{-1}$ &$10^{-3}$ \\
\vspace{-1 em}\\
\hline
\end{tabular}
\caption{\label{bella}\small Range of wavelengths of different GW experiments in units of parsec [pc]. We denote the wavelength that we will take as a reference for order of magnitude estimates by $\lambda_M$.  }
\end{center}
\end{table}

\subsection{Diffraction by compact and extended objects} \label{computation diffraction}

The metric associated to a compact massive object with $r_s=2 MG$ is the Schwarzschild metric. In Schwarzschild coordinates $(t, r, \theta, \phi)$  the metric can be written as 
\be
\bar{g}_{\mu\nu}dx^{\mu}dx^{\mu}=-(1+2\Phi)dt^2+(1+2\Phi)^{-1} dr^2+r^2d\theta^2+r^2\sin\theta^2d\phi^2\,,
\ee
with $\Phi=-GM/r$. In Lorenzian coordinates one has
\be
\bar{g}_{\mu\nu}dx^{\mu}dx^{\mu}=-\left(\frac{1-M G/(2\rho)}{1+M G/(2 \rho)}\right)^2dt^2+\left(1-\frac{M G}{2 \rho}\right)^4\left( d\rho^2+\rho^2d\theta^2+\rho^2\sin\theta^2d\phi^2\right)\,,
\ee
with
\be
r=\rho\left(1-\frac{M G}{2 \rho}\right)^2\,.
\ee
For such a spacetime the Kretschmann scalar is given by
\be
\sqrt{\bar K}=\sqrt{\bar{R}_{\mu\alpha\nu\beta}\bar{R}^{\mu\alpha\nu\beta}}=2\sqrt{3}\,\frac{r_s}{r^3}\,.
\ee
Using eq. (\ref{go}) giving the regime of validity of geometric optics, we find that wave-like effects are expected in a  region 
\be\label{com}
r_s\leq r\leq R_{\lambda}\,,
\ee
with
\be\label{Rl}
R_{\lambda}\equiv \left(2\sqrt{3}\, \lambda^2 r_s\right)^{1/3}\,.
\ee
The condition (\ref{com}) can be satisfied if $r_s\leq R_{\lambda}$, which up to factors of order unity is equivalent to $r_s<\lambda$. Note that one needs to include a factor of redshift $\lambda_{\obs}=(1+z) \lambda$  so that 
\be\label{compact con}
\frac{\lambda_{\obs}}{\text{pc}} \geq 10^{-13}\beta(1+z)\,,
\ee
where $\beta=M/M_{\odot}$. We recall that for stellar mass black holes $\beta\in [5-50]$ while for a super massive black hole $\beta\in[10^4-10^9]$. 
Choosing typical values of $\beta$ for the two cases, taking into account that most astrophysical sources are located at $z\sim1-2$ and using the values of $\lambda_{\obs}$ tested by present experiments (see table \ref{bella})  we find that for solar mass BH, diffraction effects are present in all frequency bands, while for supermassive BH diffraction is relevant only in the PTA and (partially) in the LISA bands. 

Primordial black holes have a much broader redshift distribution. For a given frequency band and average mass $\beta$, using eq. (\ref{compact con}), one can derive a condition on the redshift range where polarization is generated through diffusion.\footnote{For example, assuming that all primordial black holes have $M\sim M_{\odot}$, from eq. (\ref{compact con}) and using the average value of wavelength of different experiments (see table \ref{bella}), we find that in the LIGO, LISA and PTA band polarization effects are coming from redshift up to $z\simeq 10^3$, $z\simeq 10^6$ and $z\simeq 10^{10}$ respectively. As $\lambda_O$ scales like $\beta(1+z)$, this redshift scales like $M^{-1}$. }

For a spherically symmetric extended object with radius $R$, mass $M$ and constant density $\rho$, the gravitational potential is given by  
\be
\Phi=
\begin{dcases*}
-\frac{GM}{r}\,,&\, $r>R$\,,\\
-GM\frac{(3R^2-r^2)}{2 R^3}\,,&\, $r\leq R$\,.\\
\end{dcases*}
\ee
The corresponding metric can be written by making use of Lorenzian coordinates as
\be
g_{\mu\nu}dx^{\mu} dx^{\nu}= -e^{2\Phi(\rho)}dt^2+e^{-2 \Phi(\rho)}\left( d\rho^2+\rho^2d\theta^2+\rho^2\sin\theta^2d\phi^2\right)\,.
\ee
Note that, to first order in $\Phi$, the difference between Schwarzschild and Lorentzian coordiates can be neglected. 
At leading order in $r_s$, one has (in Schwarzschild coordinates)
\be
\bar{R}_{\mu\alpha\nu\beta}\bar{R}^{\mu\alpha\nu\beta}=
\begin{dcases*}
12\frac{r_s^2}{r^6}\,,&\, $r>R$\,,\\
15 \frac{r_s^2}{R^6}\,,&\, $r\leq R$\,.\\
\end{dcases*}
\ee
The fact that the Kretschmann scalar is discontinuous at $r=R$ is not surprising since also the density and therefore the Ricci tensor jump at this value of $r$. 
Outside the object, wave-like effects are present in a  region
\be\label{RRl}
R<r\leq R_{\lambda}\,,
\ee
with $R_{\lambda}$ defined in eq. (\ref{Rl}). Writing $R=\alpha R_{\odot}$, $M=\beta M_{\odot}$ and setting $\gamma =\beta/\alpha^3$,  
 (\ref{RRl}) can be verified if $R\leq R_{\lambda}$, i.e., if
\be\label{BBl}
\lambda_{\obs}> \frac{1}{\sqrt{\gamma}}(1+z)10^{-6}\text{pc}\,.
\ee
Inside the object, corrections to geometric optics are present for $\sqrt{15}r_s/R^3>\lambda^{-2}$ which also reproduces roughly condition (\ref{BBl}). 
Considering the values of $\gamma$ listed in table \ref{bella2}\footnote{In table \ref{bella2} we consider the average values of masses for main sequence stars, with $\beta \leq 40$. More massive main sequence stars (with mass up to $\sim 200 M_{\odot}$) exist, but  are quite rare, see e.g. \cite{2018arXiv180703821S}. For comparison, we included the range of average radii and masses of super giants, even if this is a very short lived stage of stellar evolution.} and the typical wavelengths for different experiments are given in table \ref{bella}, we find that most stars produce diffraction effects in the PTA band, in addition white dwarfs  give contributions in the LISA band while wavelengths of the LIGO band are not affected by diffraction effects from extended objects. 

\begin{table}
\begin{center}
\begin{tabular}{cccc}
Category         &    $\alpha=R/R_{\odot}$  &$\beta=M/M_{\odot}$ &  $\gamma=\beta/\alpha^3$\\
\hline\\
\vspace{-2 em}\\
Main sequence & $0.1-20$ & $0.1-40$ & $0.01-45$ \\
White dwarfs        &  $0.003-0.03$ & $0.17-1.33$ & $6\cdot 10^{4}- 5\cdot 10^{7}$ \\
\textcolor{black}{Super giants}       & $30-1500$ & $8-12$ & $10^{-9}-10^{-4}$ \\
\vspace{-1 em}\\
\hline
\end{tabular}
\caption{\label{bella2} Masses, radii and compactness of different types of stars \cite{Zombeck-book, 2010A&ARv..18...67T, 1935MNRAS..95..207C, 1990RPPh...53..837K, 2015A&A...575A..60M}.}\end{center}
\end{table}

\subsection{Integrated cross section}\label{integrated}

We have found that for both compact and extended objects, wave-like effects lead to polarization occuring in a region defined by the conditions 
(\ref{com}) and (\ref{RRl}) respectively. These conditions fix an upper bound on the impact parameter $b_{\text{max}}=R_{\lambda}$ and correspondingly a lower bound on the deflection angle 
\be
\theta_{\text{min}}=\frac{2 r_s}{b_{\text{max}}}\simeq \left(\frac{r_s}{\lambda}\right)^{2/3}\,.
\ee
Moreover, when dealing with a compact object, absorption  occurs for $b<r_s$ and this defines a maximum value for the scattering angle $\theta_{\text{max}}=2$. This bound is absent in the case of an extended object. The differential cross section (\ref{cross}) summed over polarizations  is given by eq. (\ref{cross1}) for $s=2$. 
The total cross section can be found by integrating (\ref{cross1}) over angles with $\theta\in [\theta_{\text{min}}, \theta_{\text{max}}]$. The result is
\begin{align}\label{full}
\sigma(\lambda, r_s)=\frac{2\pi}{3}(GM)^2&\left[-111\cos\theta-6  \cos(2\theta)-\cos(3\theta)-48\sin^{-2}\left(\frac{\theta}{2}\right)-384 \log\left(\sin\frac{\theta}{2}\right)\right]_{\theta_{\text{min}}}^{\theta_{\text{max}}}\,.
\end{align}
 Tor  $r_s\ll \lambda$, $\theta_{\text{min}}$ is very small and the expression in  square brackets can be approximated by the fourth term, which yields 
\be\label{sigma}
\sigma(\beta, \lambda_{\obs}, z)\simeq 10^{-7} \beta^{2/3} \left(\frac{\lambda_{\obs}}{\text{pc}}\right)^{4/3}(1+z)^{-4/3}\,\text{pc}^2\,,
\ee
where we used $\lambda_{\obs}=(1+z)\lambda$. 
This approximation is always well justified for stellar mass black holes and extended objects: the minimum wavelength we have access to is $\lambda\simeq 10^{-11} \text{pc}$ and considering $r_s(M_{\odot})\simeq 10^{-13}$ pc, we have $r_s(M_{\odot})/\lambda\leq r_s(M_{\odot})/\lambda_{\text{min}}\simeq 10^{-2}$. For supermassive black holes, we can have $\lambda \simeq r_s$ and the full expression (\ref{full}) for the integrated cross section must be used. We will however make use of (\ref{sigma}) for a first  estimate (the error is a few percent).

We now consider the case of an extended object and an impact parameter smaller than the size of the object $b<R$.  We effectively describe wave-like effects in a region (\ref{BBl}) inside the object  as a process of scattering of GW off a spherical target of radius $b$ and mass $M_b=M(b/R)^3$. The scattering angle for such a process is fixed and given by $\theta\approx 2 r_s/b$. In this case the differential cross section (\ref{cross1}) can be approximated as 
\be
\frac{d\sigma}{d\Omega}\approx 8 (GM_b)^2 \left(\frac{b}{r_s}\right)^4\,.
\ee
Writing, as usual, $R=\alpha R_{\odot}$ and $M=\beta M_{\odot}$, the expression above can be simplified to
\be
\frac{d\sigma}{d\Omega}\approx 2 \cdot 10^{-6}\, \frac{\alpha^4}{\beta^2} \left(\frac{b}{R}\right)^{10}\text{pc}^2\leq 2 \cdot 10^{-6}\,\frac{\alpha^4}{\beta^2}\text{pc}^2 \,.
\ee
The integrated cross section is (the integration over angles is simply an integration over $\phi$ and gives a factor $2\pi$) 
\be\label{sigmaR}
\sigma(\alpha, \beta, b)\simeq 10^{-5}\,\frac{\alpha^4}{\beta^2}\left(\frac{b}{R}\right)^{10}\text{pc}^2\,.
\ee
Eqs. (\ref{sigma}) and (\ref{sigmaR}) are the final results of this section.

\subsection{Optical depth}

For scattering off compact objects and extended objects with $b>R$, the integrated cross section eq. (\ref{sigma}), is redshift-dependent. For this case, the optical depths, eq. (\ref{tau}) for scattering in a matter dominated universe can be rewritten as
\be\label{taum}
\tau(z)\simeq 6\, n\, \bar{\sigma}\left(H_0\sqrt{\Omega_m}\right)^{-1}\left[(1+z)^{1/6}-1\right]\,,
\ee
where $n$ is the comoving number density of targets and $\bar{\sigma}$ is the redshift-independent part of the cross section (\ref{sigma}),   
\be
\bar{\sigma}\equiv 10^{-7} \beta^{2/3} \left(\frac{\lambda_{\obs}}{\text{pc}}\right)^{4/3}\,\text{pc}^2\,.
\ee
Analogously, for scattering taking place in a radiation dominated universe we have 
\be\label{taur}
\tau(z)\simeq 3\, n\, \bar{\sigma}\left(H_0\sqrt{\Omega_r}\right)^{-1}\left[\frac{1}{(1+z_{\text{eq}})^{1/3}}-\frac{1}{(1+z)^{1/3}}\right]\,, 
\ee
where $z_{\text{eq}}$ is the redshift at equality i.e., equal energy density in matter and radiation. 
We will make use of this last expression only when considering the case of primordial black holes. For scattering off extended objects with $b<R$, the integrated cross-section (\ref{sigmaR}) does not depend on redshift. The optical depth for this case and for $z\ll z_{\text{eq}}$ can be written as
\be\label{taum2}
\tau(z)\simeq n\, \sigma\left(H_0\sqrt{\Omega_m}\right)^{-1}\frac{2}{3}\left[(1+z)^{3/2}-1\right]\,, 
\ee
with $\sigma$ given by eq. (\ref{sigmaR}). 
In the following we compute the optical depth for scattering off different types of astrophysical objects, namely stellar mass black holes, supermassive black holes, primordial black holes and extended objects (stars).

The energy density of baryons in the observed universe is given by 
\be
\rho_B=\rho_c\Omega_B\simeq M_{\odot}\, \text{kpc}^{-3}\,.
\ee
Only about 10\% of the total baryonic matter has collapsed sufficiently to form stars and galaxies; the remaining 90\% makes up the gas in clusters and the intergalactic medium. Denoting the number of stars, stellar mass black holes (BH) and supermassive black holes (SBH) in a galaxy by $N_*$, $N_{\BH}$ and $N_{\SBH}$, we expect $N_{\SBH}=10^{-8} N_{\BH}=10^{-8}(10^{-3} N_*)$.  Then, assuming that all stars and stellar mass black holes have mass equal to the solar mass and assuming that sources are homogeneously distributed (i.e. we neglect the presence of structures in this first step), we find that the comoving density of stars is\footnote{We are making the assumption $0.1 \rho_B=M_{\odot}(n_*+n_{\BH}+\beta n_{\SBH})\approx M_{\odot} n_*$ where $\beta\in [10^{4}, 10^9]$.}
\be\label{ns}
n_*\approx \frac{\rho_* }{M_{\odot}}=0.1 \, \text{kpc}^{-3}\,,
\ee
and for stellar mass black holes and supermassive black holes, respectively  
\be
n_{\BH}\simeq 10^{-4}\, \text{kpc}^{-3}\,, 
\ee
\be
n_{\SBH}\simeq 10^{-12}\, \text{kpc}^{-3}\,. 
\ee
%{\color{red}IS THIS CORRECT? HOW DO WE ACCOUNT FOR THE FACT THAT STARS ARE CLUMPED IN GALAXIES WHERE THE DENSITY IS MUCH, MUCH HIGHER?}
Using eq. (\ref{sigma}) in eq. (\ref{taum}) we obtain
\be
\tau_{\BH}(z, \lambda_{\obs})\simeq10^{-9}\left(\frac{\lambda_{\obs}}{\text{pc}}\right)^{4/3}\left[(1+z)^{1/6}-1\right]\,.
\ee
Analogously, for supermassive black holes
\be
\tau_{\SBH}(z, \lambda_{\obs}, \beta)\simeq10^{-17}\beta^{2/3}\left(\frac{\lambda_{\obs}}{\text{pc}}\right)^{4/3}\left[(1+z)^{1/6}-1\right]\,,
\ee
with $\beta\in [10^4, 10^9]$.

Next, we consider  the possibility that primordial black holes (PBH) represent a fraction $q$ of the dark matter energy density, i.e.
\be
\Omega_{\PBH}=q \Omega_{\CDM}\,,
\ee
with $q \in [0, 1]$. The comoving number density of primordial black holes can be estimated as
\be
n_{\PBH}=\rho_c \frac{\Omega_{\CDM}}{M_{\PBH}}q \,.
\ee
Writing $M_{\PBH}=\beta M_{\odot}$ with $\beta <10^3$ this yields
\be\label{nP}
n_{\PBH}\simeq \frac{q}{\beta}\, (\text{kpc})^{-3}\,.
\ee
PBH have a broad redshift distribution, which extends up to very high redshift, see e.g. \cite{Sasaki:2018dmp} for a recent review. It follows that a cosmological background of GW radiation (produced during or after inflation) can scatter in a broad redshift range. 
%Comparing the functional dependence of (\ref{taum}) and (\ref{taur}), we expect the dominant contribution to the visibility function to come from scattering at high redshift. 
Inserting Eqs. (\ref{sigma}) and (\ref{nP}) in (\ref{taur}), we find
%\be
%\tau^{\text{cosmo}}_{\PBH}(z, \lambda_{\obs}, \beta)\simeq\frac{q}{\beta^{1/3}}10^{-4}\left(\frac{\lambda_{\obs}}{\text{pc}}\right)^{4/3}\left[\frac{1}{(1+z_{\text{eq}})^{1/3}}-\frac{1}{(1+z)^{1/3}}\right]\,.
%\ee
\be
\tau^{\text{cosmo}}_{\PBH}(z, \lambda_{\obs}, \beta)\simeq\frac{q}{\beta^{1/3}}10^{-4}\left(\frac{\lambda_{\obs}}{\text{pc}}\right)^{4/3}\frac{1}{(1+z_{\text{eq}})^{1/3}}\,.
\ee
%\rut{[NOTE: I don't understand this. In the rd universe you get a result roughly $\propto (1+z_{\text{eq}})^{-1/3}/\sqrt{\Omega_r}$ while in the md universe it is $\propto (1+z_{\text{eq}})^{1/6}/\sqrt{\Omega_m}$. Now  $\Omega_r/\Omega_m =1/(1+z_{\text{eq}})$ hence both contributions are roughly the same???]}\\
For an astrophysical background of GW, we use (\ref{taum}) to find
\be
\tau^{\text{astro}}_{\PBH}(z, \lambda_{\obs}, \beta)\simeq\frac{q}{\beta^{1/3}}10^{-5}\left(\frac{\lambda_{\obs}}{\text{pc}}\right)^{4/3}\left[(1+z)^{1/6}-1\right]\,.
\ee

%\subsection{Optical dephth for extended objects}

We turn now to the case of GW scattering off a distribution of stars.  The condition to have diffusion is given by eq. (\ref{BBl}). We use the results in tables \ref{bella} and \ref{bella2}: diffraction occurs for all the physical values of $\gamma$ in the PTA band, in the LISA band a contribution comes only from scattering off white dwarfs while no effect is present in the LIGO band. Assuming that all the stars have have Solar mass, the comoving density of stars is given by eq. (\ref{ns}). Then, in the PTA, replacing (\ref{ns}) and (\ref{sigma}) in (\ref{taum}) and choosing $\beta=1=\alpha$, we find
\be
\tau^{\text{PTA}}_{*\text{out}}(z, \lambda_{\obs})\simeq 10^{-6}\left(\frac{\lambda_{\obs}}{\text{pc}}\right)^{4/3}\left[(1+z)^{1/6}-1\right]\,, 
\ee
 for scattering with impact parameter bigger than the size of the object. 
For scattering off extended objects with impact parameter smaller than the object size, the (effective) cross section describing such a process is given in eq. (\ref{sigmaR}).  Then, in the PTA band, assuming $\beta=1=\alpha$ and plugging (\ref{sigmaR}) and (\ref{ns}) in (\ref{taum2}), we have
\be
\tau^{\text{PTA}}_{*\text{in}}(z, b)\simeq 10^{-5}\left(\frac{b}{R_{\odot}}\right)^{10} \left[(1+z)^{3/2}-1\right]\,.
\ee
Assuming the impact parameter $b$ to be distributed uniformly  in $[0, R_{\odot}]$ and choosing, for the estimate, the average value $\langle b\rangle=R_{\odot}/2 $, we find
\be
\tau^{\text{PTA}}_{*\text{in}}(z)\simeq 10^{-8} \left[(1+z)^{3/2}-1\right]\,.
\ee
For the LISA band, denoting as $f_{\text{wd}}$ the fraction of stars in white dwarfs, we find
\be
\tau^{\text{LISA}}_{*\text{out}} (z, \lambda_{\obs})\simeq 10^{-6}\,f_{\text{wd}}\left(\frac{\lambda_{\obs}}{\text{pc}}\right)^{4/3}\left[(1+z)^{1/6}-1\right]\,,
\ee
and\footnote{The local (mid-plan) mass density of white dwarfs is $\rho_{\text{wd}}\simeq 0.0065 M_{\odot} \text{pc}^{-3}$, see e.g. \cite{Holberg:2001in}. Considering that the average mass of stars is $\sim 0.5 M_{\odot}$ and comparing to the local density of stars, see e.g. \cite{2017MNRAS.470.1360B}, one obtains the local value $f_{\text{wd}}\sim 0.05$.}
\be
\tau^{\text{LISA}}_{*\text{in}}(z)\simeq 10^{-8}\,f_{\text{wd}}\, \left[(1+z)^{3/2}-1\right]\,.
\ee

We emphasize that here we  assume that astrophysical sources have an isotropic distribution.  As we will see in the next section \ref{coherent}, the average distance between objects in a structure is such that the multi-scattering of gravitational waves off massive structures can be considered as incoherent and the total polarization generated through diffusion is the linear sum of polarization created off different scatterings. It follows that, as long as we are interested in deriving an estimate of the total polarization generated (averaged over directions at the observer position), neglecting the presence of structures is  well justified.

\subsection{Coherent and incoherent scattering}\label{coherent}

Let us consider a distribution of compact objects/stars, with density $n$ and mean distance $d=n^{-1/3}$ (i.e. $d$ is the average distance between objects) and masses $M_i$. If we have a wave with wavelength $\lambda$ incident on this distribution, depending on the relative size of $\lambda$ and $d$, the process has to be treated as either coherent or incoherent. To be specific we have the that
\begin{itemize}
\item if $\lambda \gg d$ the scattering is coherent; 
\item  if $\lambda \ll d$ the scattering is incoherent.
\end{itemize}
If the scattering is coherent, then the multi-scattering process can be treated as a single scattering off a target with mass $M_{\text{tot}}=\sum_i M_i$. The total cross section is given by
\be
\frac{d\sigma}{d\Omega}\propto \left(\sum_i M_i\right)^2 G^2 [\dots]\,,
\ee
where $[\dots]$ stays for some angular structure. If the scattering is incoherent, the total cross section is given by the linear sum  of single cross sections
\be
\frac{d\sigma}{d\Omega}\propto \sum_i \left(M_i G\right)^2 [\dots]\,. 
\ee

Let us consider a GW passing through a globular cluster. The number of stars in a globular cluster is $N_*\simeq 10^{4}-10^5$ and the (typical) radius of the cluster $R_{\text{GC}}\simeq 10$ pc. The average distance between objects  is then given by $d_{\text{GC}}=n_{\text{GC}}^{-1/3}\simeq(V_{\text{GC}}/N_*)^{1/3}\simeq 1$ pc (where $V_{\text{GC}}$ is the volume of the cluster). Comparing with the wavelength we have access to observationally (see table \ref{bella}) we always have $d_{\text{GC}}>\lambda_{\obs}>\lambda$: the multi-scattering  process can be treated as incoherent scattering (some coherence could be present in the low frequency part of the PTA range). If we repeat the same reasoning for a galaxy we get $d>d_{\text{GC}}$ (objects are less densely distributed) and the same conclusion holds, except maybe close to the galactic center where scattering however is dominated by the central super massive black hole.

\subsection{The total amount of polarization.}\label{visibility 3}

For a given wavelength, the total amount of polarization produced by diffusion is given by the sum of the polarization produced from  scattering off different types of massive structures. We recall that the condition to have diffusion effects in the vicinity of a compact object is given by eq. (\ref{compact con}) while the analogous condition for an extended object is (\ref{BBl}). Using the results in tables \ref{bella} and \ref{bella2}, we find that in different frequency bands, different objects are causing diffraction effects which lead to polarization 
\begin{itemize}
\item PTA band: solar mass BH, supermassive black holes, any type of star; 
\item  LISA band: solar mass BH, (some) (super)massive black hole, white dwarfs; 
\item LIGO band: solar mass BH.
\end{itemize}
The condition to have scattering off primordial black holes is more subtle since PBH have a broad redshift distribution. Considering a vanilla model where all PBH have the same mass $M=\beta M_{\odot}$, for a given observed frequency, the condition (\ref{compact con}) gives  the maximum redshift at which we can have diffusion and polarization generation. In particular, all solar mass PBH (or lighter) up to $z=z_{\text{eq}}$ act as scattering centers and produce polarization of the background in the entire frequency range of present and planned GW observatories.

We can now work out an estimate for the polarization fraction (\ref{QQQ}) for the LIGO, LISA and PTA band. To do so, we introduce a number of simplifications. For the three cases, we choose $\lambda_M$ from Table \ref{bella}. Moreover, since most astrophysical sources are expected to be located around redshift $z=1-2$ (see e.g. \cite{Madau:2014bja}) we evaluate the visibility function at $z=1.5$. If primordial black holes represent a fraction $q$ of the total dark matter component, we distinguish two cases: scattering taking place in matter domination and in radiation domination. Obviously, only a background of cosmological origin can undergo the latter type of process. In this first approximation, we assume that the optical depth for scattering off PBH at early times is redshift independent.\footnote{At earlier times, the optical depth has a mild redshift dependence, see eq. (\ref{taur}) which we neglect.} At late times we assume the baryonic matter distribution to follow the distribution of primordial black holes and we consider most of these targets to be located around $z=1.5$. 
%We recall that $f_{\text{wd}}$ we denote the fraction of stars which is white dwarfs. 
Our results are summarized in table \ref{bella3}.

\begin{table}
\begin{center}
\begin{tabular}{cccc}
        &    PTA  ($\lambda_M=10^{-3}$) &LISA ($\lambda_M=10^{-7}$)& LIGO  ($\lambda_M=10^{-10}$)\\
\hline\\
$\tau_{\BH}$ & $10^{-14}$ & $10^{-19}$ & $10^{-23}$ \\
$\tau_{\SBH}$       &  $\beta^{2/3}10^{-22}$ & $\beta^{2/3}10^{-27}$& $0$ \\
$\tau_{*, \text{in}}$      & $10^{-8}$ & $ f_{\text{wd}}\,10^{-8}$ & $0$ \\
$\tau_{*, \text{out}}$ & $10^{-11}$ & $ f_{\text{wd}}\,10^{-16}$ & $0$ \\
$\tau_{\PBH, \text{astro}}$& $q \beta^{-1/3}\,10^{-10}$ & $q\, \beta^{-1/3}\,10^{-15}$ & $q\,\beta^{-1/3}\,10^{-19}$ \\
$\tau_{\PBH, \text{cosmo}}$& $q \, \beta^{-1/3}\,10^{-9}$ & $q\, \beta^{-1/3}\,10^{-14}$&$q\, \beta^{-1/3}\,10^{-18}$  \\
\\
\hline
\end{tabular}
\caption{\label{bella3}Optical depth for scattering off different types of massive structures.}
\end{center}
\end{table}

The total amount of polarization is given by the sum of the polarization generated by different scatterings, i.e. the total parameter  (\ref{QQQ}) is given by
% {\color{red} WHAT DOES i STAND FOR AS A LABEL? TYPE OF SCATTERERS, ACTUAL SCATTERERS?}
\be
P=\sum_iP_i\simeq \frac{\delta \Omega_{GW}}{\bar{\Omega}_{GW}}\sum_i\tau_i\,, 
\ee
where $i$ labels different types of scatterers. In the last equality we have used that $\tau\ll 1$ and $\delta\Omega_{GW}/\bar{\Omega}_{GW}$ is the typical amount of anisotropies of a given component of the background. An estimate of this quantity as a function of redshift for a cosmological and astrophysical background can be found in appendix \ref{new}. For a cosmological background, in any frequency range, $\delta\Omega_{GW}/\bar{\Omega}_{GW}\sim 10^{-5}$. For an astrophysical background the result depends on frequency, on the astrophysical sources we consider and on redshift. In the LIGO band the background is dominated by the contribution from black hole mergers and at $z=1.5$ where we assume most of the sources producing scattering to be located, we have $\delta\Omega_{GW}/\bar{\Omega}_{GW}\sim 10^{-3}$. If we extrapolate the LIGO estimate of the astrophysical background to all frequencies we then have  
\be
\frac{\delta \Omega_{GW}}{\bar{\Omega}_{GW}}\simeq 
\begin{dcases*}
10^{-3}\,,&\, for astrophysical background\,,\\
10^{-5}\,,&\, for cosmological background\,.\\
\end{dcases*}
\ee

In the LIGO band the dominant contribution to polarization comes from diffusion off PBH and stellar mass black holes. The total amount of polarization generated is given by 
\be\label{LIGO1}
P_{\text{LIGO}}\simeq 
\begin{dcases*}
10^{-22}\left(q\beta^{-1/3}+10^{-4}\right)\,,&\, for astrophysical background\,,\\
10^{-23}\left(q\beta^{-1/3}+10^{-5}\right)\,,&\, for cosmological background\,.\\
\end{dcases*}
\ee
In the LISA band, the dominant contribution to polarization generation comes from scattering off white dwarfs and the amount of polarization generated is given by 
\be\label{LISA1}
P_{\text{LISA}}\simeq f_{\text{wd}} 10^{-11}\,,
\ee
for an astrophysical background while for a cosmological background, the result is suppressed by an additional factor $10^{-2}$. In the PTA band, the processes generated more efficiently polarization are scattering offf PBH and extended objects. One finds 
\be\label{PTA1}
P_{\text{PTA}}\simeq 
\begin{dcases*}
10^{-11}\left(1+0.01 \,q\beta^{-1/3}\right)\,,&\, for astrophysical background\,,\\
10^{-13}\left(1+0.1\, q\beta^{-1/3}\right)\,,&\, for cosmological background\,.\\
\end{dcases*}
\ee
To get a first order of magnitude estimate of the effect, we consider the simplest scenario of PBH with monochromatic mass distribution $M=M_{\odot}$ and $q=0.5$, still allowed by current observational bounds, and we extrapole the local value $f_{\text{wd}}=0.05$  at extragalactic scales. The results are listed in table \ref{finale tabella}.  
\begin{table}
\begin{center}
\begin{tabular}{cccc}
        &    PTA  ($\lambda_M=10^{-3}$) &LISA ($\lambda_M=10^{-7}$)& LIGO  ($\lambda_M=10^{-10}$)\\
\hline\\
\vspace{-2 em}\\
$P_{\text{astro}}$ & $10^{-11}$ & $10^{-12}$ & $10^{-23}$ \\
$P_{\text{cosmo}}$       &  $10^{-13}$ & $10^{-14}$& $10^{-24}$
\\
\vspace{-1 em}\\
\hline
\end{tabular}
\caption{\label{finale tabella}Polarization of an astrophysical and cosmological generated through diffusion off massive structures in different frequency bands. The estimate has been obtained under the assumptions of section \ref{visibility 3}: PBH with monochromatic mass distribution and $\beta=2 q=1$ and average fraction of white dwarfs equal to the local one. }
\end{center}
\end{table}

We emphasize that the estimates (\ref{LIGO1}), (\ref{LISA1}) and (\ref{PTA1}) for a cosmological background are robust since the prediction $\delta\Omega_{GW}/\bar{\Omega}_{GW}\sim 10^{-5}$ does not depend on astrophysical complications and on frequency (see appendix \ref{new}). For an astrophysical background, we are extrapolating the prediction for $\delta\Omega_{GW}/\bar{\Omega}_{GW}$ valid in the LIGO band to lower frequencies. To derive a more realistic estimate of $P$ for an astrophysical background in the PTA and LISA band, one has to repeat the analysis of \cite{Cusin:2018rsq} valid for the LIGO frequency band, and work out the amplitude of anisotropies at lower frequency, including contributions from different astrophysical sources. This analysis will be presented  in \cite{preparation}.

\section{Conclusions}\label{conclusion}

In this work, we have discussed the production of polarization of a stochastic GW background from diffusion by extended and compact astrophysical objects. We have provided a framework which can be used to compute the polarization tensor of a given component of the background, in any frequency range. The main ingredients of our approach are the following.  We compute the integrated cross section for scattering 
off a given massive structure. Since the geometric optics approximation breaks down in a region of radius $R_{\lambda}$ around the scatterer, the impact parameter has an upper bound $b_{\max}=R_{\lambda}$ and with $b_{\max}$, the integrated cross section for a compact (extended) object depends on the mass (on the compactness) of the object and on the wavelength of the GW. As a consequence, the optical depth depends on the abundance of targets, on the properties of the target and on the wavelength. We have discussed that, for the wavelength range we  have access to in present and planned GW experiments, the average separation between scattering centers is much larger than the wavelength, hence multi-scattering can be studied as an incoherent sum of scattering processes. To compute the total polarization which the GW background acquires due to interaction with structures, it is therefore sufficient to sum the polarization generated by different types of scatterings. 

In our framework, the back reaction of polarization on intensity is systematically discarded. More precisely, we compute the intensity of the GW background measured by a comoving observer at a given redshift and from a given direction, using the geometric optics approximation and we use this result to compute the polarization created  from a scattering event. The geometric optics approximation is not suitable to describe the creation of polarization by interaction with matter:  wave like effects represent a first order correction to the intensity while they give a zero order contribution to polarization.  Since the polarization created by scattering is very suppressed with respect to the intensity, see section \ref{visibility 1},  neglecting the back reaction of polarization on intensity is  well justified. 

We have also assumed that the flux of radiation at emission is unpolarized. As already discussed, this is a very good assumption for the case of a cosmological background, for which we expect that the two polarizations are generated with equal probability.  
If the initial  (primary) polarization is sufficiently small, it can be simply added to the secondary one created during the propagation to the observer position, neglecting back reaction. A more rigorous approach consists in writing a set of Boltzmann equations for intensity and polarization, with a  collision term that is coupling the system. Writing such a system is rather straightforward : one needs to use the results of appendix \ref{Bol} and derive the angular dependence of the collision term from the angular dependence of the differential cross section, eq. (\ref{cross}).\footnote{However, an additional complication with respect to the case of the CMB comes from the fact that, the differential cross section presents a Rutherford-like small angle behavior with inverse powers of $(1-\cos\theta)$. Therefore, the angular dependence of the collision term would involve an infinite expansion of Legendre polynomials of the scattering angle, $\cos\theta$.}

In section \ref{visibility 1} we have derived an order of magnitude estimate for the polarization in different frequency bands for both astrophysical and cosmological GW backgrounds.  In the LIGO band the dominant contribution to the creation of polarization comes from scattering off stellar mass black holes or PBH if they exist; in the LISA band scattering off white dwarfs dominates while in the PTA band from scattering off stars (or PBH if they exist) is most important.  For an astrophysical background in the LIGO band, in the scenario in which primordial black holes have a monochromatic mass distribution with $M=M_{\odot}$ and constitute half of the total amount of dark matter\footnote{In Ref.\,\cite{Zumalacarregui:2017qqd}, bounds on the abundance of compact objects from gravitational lensing of type Ia supernovae are derived: compact objects represent less than $\sim 40\%$ of the total matter content in the universe, at 95\% confidence-level, thus excluding a scenario of all dark matter made up by primordial black holes. Ref.\,\cite{Garcia-Bellido:2017imq} criticizes some aspects of Ref.\,\cite{Zumalacarregui:2017qqd} and shows  that all-PBH dark matter scenario in the LIGO band is compatible with SN lensing constraints. The criticism  is then addressed in the published version of \cite{Zumalacarregui:2017qqd}, showing the validity of the constraints previously derived. However, in the case of a monochromatic mass distribution and no-clustering assumption EROS bounds \cite{Tisserand:2006zx} hold and disfavor a value $q=1$ in this mass range.}, the amount of polarization created is suppressed with respect to (energy density) anisotropies by a factor $10^{-20}$ and $10^{-19}$ for an astrophysical and cosmological background, respectively.  An enhancement of this result can be obtained in the case in which the mass distribution of PBH is not monochromatic and /or peaked at $M\ll M_{\odot}$. Observational bounds on the relative abundance of PBH for $M\ll M_{\odot}$  comes mainly from the EROS microlensing survey \cite{Tisserand:2006zx} and from the cosmic microwave background temperature anisotropies and spectral distortions \cite{Ricotti:2007au}. However, these bounds have recently been re-investigated: EROS limits can easily be evaded \cite{Hawkins:2011qz, Green:2017qoa, Garcia-Bellido:2017xvr}, e.g. when considering more realistic dark matter distributions in the galaxy or if most PBH are regrouped in micro-clusters.  On the other hand, Planck limits on PBH abundances have been shown to be very sensitive to the PBH velocity with respect to baryons, while there is no relevant constraint from CMB spectral distortions \cite{Poulin:2017bwe, Ali-Haimoud:2016mbv}. This reopened the low-mass window for PBH as dark matter candidate.

The generation of polarization is considerably enhanced at lower frequencies: in the LISA band the suppression of polarization with respect to anisotropies if $\sim 10^{-9}$ for both an astrophysical and cosmological background while for PTA the effect is enhanced by a factor 100 and the suppression of polarization with respect to anisotropies is of order $10^{-7}$. The estimate in the LISA band is derived assuming that the average fraction of stars in white dwarfs at extragalactic scales is the local one. The results for PTA can be further enhanced in a  scenario in which PBH have a mass distribution peaked at $M\ll M_{\odot}$. Just to have a term of comparison, for the CMB, the amount of polarization generated is suppressed by a factor $\sim 10^{-2}$  with respect to temperature fluctuations. 

Our results are particularly interesting for a cosmological GW background. For this case the primary polarization vanishes and an upper bound on polarization can set upper bounds on the abundance of unresolved objects in the universe. For example, setting an upper bound on polarization in the LISA band, would set a bound on the extragalactic value of $f_{\text{wd}}$, see eq. (\ref{LISA1}), which we have set equal to its local value  in our estimates. Analogously, observing in the PTA band, an enhancement of polarization may provide information on the PHB mass distribution and abundance in the window $M\ll M_{\odot}$. 

We emphasize that the aim of the estimates in section \ref{visibility 1}  is to get an idea of the  size of the effect and of the kind of information that could in principle be extracted if polarization is measured or if stringent upper bounds are obtained.  In particular, in section \ref{visibility 1} we have introduced the simplifying hypothesis that the distribution of targets is isotropic. As explained, this assumption is consistent as long as we are interested in obtaining an estimate of the average amount of polarization, integrated over directions at the detector position. A  direction dependent visibility function can be introduced as in section \ref{end} and the more refined setting of section \ref{general} and \ref{analytical} can be used to derive more precise results. In particular, since astrophysical objects acting as targets are embedded in  galaxies which in turn belong to clusters etc., we expect polarization to have a pronounced directional dependence. Therefore, even if the averaged degree of polarization is very small, it may be much stronger and possibly detectable in the direction of compact structures or when correlated with matter over-densities. Detailed evaluations of experimental possibilities are left for future work. Moreover, as explained in section \ref{analytical}, GW polarization is a stochastic quantity which can be characterized statistically in terms of its angular power spectrum and its cross-correlation with intensity, in full analogy with the CMB polarization. From an observational point of view, methods are already available to reconstruct a polarization map of the sky, see e.g. the review \cite{Romano:2016dpx}. The algorithm proposed in \cite{Renzini:2018vkx} to reconstruct a sky map of intensity can also be generalized to polarization. Comparing a sky map of polarization with theoretical predictions can be extremely interesting to reconstruct a chart of the \emph{invisible} universe. For example, measuring an overproduction of polarization in a given direction would be an indication of the presence of a dense structure, e.g.  a cluster (resolved or unresolved) in that direction.

\subsection*{Acknowledgements}
We thank Cyril Pitrou and Jean-Philippe Uzan for valuable comments and discussions during various stages of this work and for their help in the derivation of the emissivity function in the Boltzman approach. We are grateful to Macarena Lagos and Bernard Whiting for very useful discussions. \textcolor{black}{We thank Michele Maggiore for an inspiring conversation at the early stage of this work}. Finally, we thank Irina Dvorkin for comments on the astrophysical part of our study, Pierre Fleury and Davide Racco for discussions and references on primordial black holes and Enrico Barausse, Laura Bernard, Luc Blanchet and Guillaume Faye for useful references on GW diffusion. G.C. acknowledges  financial support from
ERC Grant No:  693024 and Beecroft Trust, R.D. acknowledges 
funding from the Swiss National Science Foundation. P.G.F. acknowledges support from Leverhulme, STFC, BIPAC  and  the  ERC.  

\newpage

\newpage

\appendix

\section{Formal results on polarization tensors}\label{basis}

 We work in transverse traceless  gauge (TT). % defined by eq. (\ref{TT}).  
 Using a plane wave expansion we can write a generic GW in TT gauge as 
\be
h^{TT}_{ij}(t, \bx)=\int df  \int d^2\bn\,\tilde{h}_{ij}(f, \bn)  e^{-2\pi if(t-\bn \cdot \bx/c)}\,,
\ee
and the Fourier components of the metric perturbation $h_{ij}(t, \bx)$ can be expanded in terms of either the linear polarization basis tensors 
\be
\tilde{h}_{ij}(f, \bn)=\tilde{h}_+(f, \bn) e_{ij}^+(\bn)+\tilde{h}_{\times}(f, \bn) e_{ij}^{\times}(\bn)\,,
\ee
or the circular polarization basis tensors
\be
\tilde{h}_{ij}(f, \bn)=\tilde{h}_R(f, \bn) e_{ij}^R(\bn)+\tilde{h}_{L}(f, \bn) e_{ij}^{L}(\bn)\,, 
\ee
where the circular polarization basis is defined in eqs. (\ref{RL}) and (\ref{LR}).  
%\bees\label{RL}
%e^R_{ij}&=&\frac{1}{\sqrt{2}}\left(e_{ij}^{+}+i\, e_{ij}^{\times}\right)\,,\\
%e^L_{ij}&=&\frac{1}{\sqrt{2}}\left(e_{ij}^{+}-i\, e_{ij}^{\times}\right)\,,
%\ees
The expansion coefficients $\tilde{h}_R, \tilde{h}_L$ are related to $\tilde{h}_+, \tilde{h}_{\times}$ through
\bees\label{lov}
\tilde{h}_R&=&\frac{1}{\sqrt{2}} (h_+-i h_{\times})\,,\\
\tilde{h}_L&=&\frac{1}{\sqrt{2}} (h_+ +i h_{\times})\,.\label{lov1}
\ees
Summarizing, we write 
\be\label{1}
h^{TT}_{ij}(t, \bx)=\sum_{A} \int df\int d^2\bn\, \tilde{h}_A(f, \bn) e_{ij}^A (\bn) e^{-2\pi if(t-\bn \cdot \bx/c)}\,,
\ee
where the sum is over the two tensors of the polarization basis, i.e. $A=(+, \times)$ if we are using the linear polarization basis and  $A=(R, L)$ if we are using the circular polarization basis.

\subsection{General properties of polarization tensors}

 The polarization tensors $e_{ij}^A(\bn)$ satisfy
\be\label{delta}
e^{A}_{ij}(\bn) e^{B\,ij*}(\bn)=2\delta^{AB}\,.
\ee
Under a rotation of an angle $\psi$ in the plane orthogonal to the $\bn$ direction, the polarization basis $(e_{ij}^+, e_{ij}^{\times})$ transforms as
\bees
e_{ij}^+(\bn, \psi)&=&\cos2\psi\,e^{+}_{ij}(\bn)+\sin2\psi\,e^{\times}_{ij}(\bn)\,,\\
e_{ij}^{\times}(\bn, \psi)&=&-\sin2\psi\,e^{+}_{ij}(\bn)+\cos2\psi\,e^{\times}_{ij}(\bn)\,.
\ees
while the  basis $(e_{ij}^R, e_{ij}^{L})$ transforms as
\bees\label{Rr}
e_{ij}^R(\bn, \psi)&=&e^{-i2\psi} e_{ij}^R(\bn)\,,\\
e_{ij}^L(\bn, \psi)&=&e^{i2\psi} e_{ij}^L(\bn)\,.\label{Ll}
\ees
The tensors $e_{ij}^R$ and $e_{ij}^L$ corresponds to right and left circularly polarized waves.

Let us construct the basis of polarization tensors $(e^{\times}_{ij}(\bn), e^{+}_{ij}(\bn))$. We can choose an orthonormal basis in the plane normal to the direction of propagation $\bn$, i.e. $\bu(\bn)$ and $\bv(\bn)$. We define
\begin{align}\label{defer}
e_{ij}^+(\bn)&=u_i(\bn)u_j(\bn)-v_i(\bn)v_j(\bn)\,,\\
e_{ij}^{\times}(\bn)&=u_i(\bn)v_j(\bn)+v_i(\bn)u_j(\bn)\,. 
\end{align}
Using the fact that $\bv$ and $\bu$ are orthonomal, one can  verify that eq. (\ref{delta}) is satisfied. We compute the contraction between polarization vectors relative to different directions of propagation. We use the shortcut  notation $e_{ij}^A(\bn')=e_{ij}'^A$. We have
\begin{align}
e_{ij}'^{+}e^{ij+}&=(\bu\cdot \bu')^2+(\bv\cdot \bv')^2-(\bv\cdot \bu')^2-(\bu\cdot \bv')^2\,,\\
e_{ij}'^{\times}e^{ij\times}&=2(\bu\cdot \bu')(\bv\cdot \bv')+2(\bv\cdot \bu')(\bu\cdot \bv')\,,\\
e_{ij}'^{\times}e^{ij+}&=2(\bu\cdot \bu')(\bu\cdot \bv')-2(\bv\cdot \bv')(\bv\cdot \bu')\,,\\
e_{ij}'^{+}e^{ij\times}&=2(\bu\cdot \bu')(\bv\cdot \bu')-2(\bv\cdot \bv')(\bu\cdot \bv')\,.
\end{align}
One can verify that\footnote{One way to verify this result is to pick up a specific choice for $\bn$, $\bn'$ and for the vectors of the polarization basis, expressing the final result in a coordinate independent way. For example we can choose $\bn=(\sin\theta, \cos\theta, 0)$, $\bn'=(0, 1, 0)$ and $\bu'=\bee_x$, $\bv'=\bee_z$ and  $\bu=(\cos\theta, -\sin\theta, 0)$, $\bv=(0,0,1)$. One finds $e_{ij}'^{+}e^{ij+}=1+\cos\theta^2$, $e_{ij}'^{\times}e^{ij\times}=2\cos\theta$ and the mixed terms are vanishing. We then simply insert $\cos\theta=\bn\cdot \bn'$.}
\be
\sum_{AB} |e_{ij}^{A}(\bn')e^{ijB}(\bn)|^2=1+(\bn\cdot \bn')^4+6(\bn\cdot \bn')^2\,.
\ee

\newpage

\section{Boltzman approach}\label{Bol}

We propose an alternative derivation of the results of \cite{Cusin:2017fwz, Cusin:2017mjm} using a Boltzmann approach.\footnote{See also \cite{Contaldi:2016koz} for a first attempt to derive anisotropies from a Boltzmann approach. }% {\color{red} we need to make sure we cite all his papers on this}.}  
We will see that the geometric optics approximation used in \cite{Cusin:2017fwz, Cusin:2017mjm} corresponds to the zero-collision hypothesis in the Boltzmann treatment. We also derive result for the anisotropies of a cosmological background of radiation. 

We introduce a distribution function of gravitons $f(x^{\mu}, k^{\mu})$, satisfying the following Boltzman equation
\be\label{LL}
\mathcal{L}[f]=\mathcal{E}[\lambda]+\mathcal{C}[f]\,,
\ee
where $\mathcal{L}\equiv d/d\eta$ is the Liouville operator and $\mathcal{E}$, $\mathcal{C}$ and emissivity term and collision term respectively. We denote by $\lambda$ the affine parameter along the geodesic with tangent vector $k^{\mu}$ so that $d\eta/d\lambda=k^0=\omega$. In this appendix we denote the frequency by $\omega$, to avoid confusion with the distribution function. 

We introduce two reference frames, a cosmological reference frame and a reference frame tied to baryons (and galaxies under the assumption that the galaxy velocity is unbiased). We then have
\be
e^{\mu}_{\,\,a}=\left(u^{\mu}\,, e^{\mu}_{\,\,i}\right)\,, \qquad \left(e^{\mu}_{\,\,a}\right)_{\Gal}=\left(u^{\mu}\,, e^{\mu}_{\,\,i}\right)_{\Gal}\,,
\ee
which are related by a boost of velocity $\bv$ as
\be
e^{\mu}_{\,\,a}=\Lambda^{\mu}_{\,\,\,\nu} (e^{\nu}_{\,\,a})_{\Gal}\,,
\ee
see appendix G of \cite{Cusin:2016kqx} for details on how physical quantities transform under this boost. If we have a graviton with 4-momentum $(k^{\mu}) =\om(1,n^i)$, its energy and direction measured in the two frames are related as
\begin{align}
\omega_{\Gal}&=-\left(u_{\mu}k^{\mu}\right)_{\Gal}=\omega(1-\bn\cdot \bv)\,,\\
n^i_{\Gal}&=(1+\bn\cdot \bv)n^i\,,
\end{align}
to first order there is no aberration.
The distribution function does not transform and one has
\be
f(\eta, \bx, \omega, \bn)=f_{\Gal}(\eta,  \bx, \omega_{\Gal}, \bn_{\Gal})\,.
\ee
From now on we therefore omit the subscript $\Gal$ on the distribution function.  We define (in the frame of the comoving observer)\footnote{For practical purposes, we are absorbing a factor $1/(4\pi)$ in the definition of $f$. To make contact with standard conventions, we will replace $\Omega_{GW} \rightarrow 4\pi \Omega_{GW}$ in the final result.}
\be
\Omega_{GW}(\eta, \bx, \bn_{\Gal})\equiv\frac{1}{\rho_c} \int d\omega_{\Gal}\, \omega_{\Gal}^3\, f(\eta, \bx, \omega_{\Gal}, \bn_{\Gal})\,. 
\ee

The Liouville operator can be written as
\be\label{qua}
\mathcal{L}[\lambda]=\frac{d\eta}{d\lambda}\left[\frac{\partial f}{\partial \eta}+n^i\partial_i f+\frac{\partial f}{\partial \log \omega_{\Gal}}\frac{\partial \log \omega_{\Gal}}{\partial \eta}\right]\,.
\ee
%\textcolor{magenta}{where $\lambda$ is an affine parameter along the geodesic with tangent vector $k^{\mu}$ so that $d\eta/d\lambda=k^0$. }
We observe that since we will be interested in quantities up to first order in perturbations, we can neglect all aberration effects in eq. (\ref{qua}), setting $\bn=\bn_{\Gal}$.  Furthermore, we have assumed that $f$ has no intrinsic direction  dependence so that the latter only enters via the dependence of $\omega_G$ on $\bn$ via 
\be
\frac{d\log \omega_{\Gal}}{d\eta}=\frac{d\log \omega}{d\eta}-n^i\frac{dv^i}{d\eta}\,.%=-\mathcal{H}-n^i\partial_i\phi+\dot{\psi}-n^i\frac{dv^i}{d\eta}\,.
\ee

\subsection{Astrophysical background}

We neglect the collision term in eq. (\ref{LL}). We then have
\be\label{LLL}
\frac{df}{d\tau}=\frac{d\lambda}{d\tau}\mathcal{E}[\lambda]\,,
\ee
where $\tau$ is the proper time of the observer.  
%\rut{[NOTE: what? a massless particle has no proper time. Should $\tau$ not rather be the proper time of the emitter or cosmic time?]}\\
We recall that $\mathcal{E}[\lambda]$=(number gravitons produced)/(units of $\lambda$). It follows that the quantity on the right hand side of (\ref{LLL}) is proportional to the number of gravitons produced per units of $\tau$. We observe that 
\be
\frac{d \rho_{GW}}{d\tau}=\int d\omega_{\Gal}\, \omega_{\Gal}^3\, \frac{df}{d\tau}=\int d\omega_{\Gal}\, \omega_{\Gal}^3\, \frac{d\lambda}{d\tau} \mathcal{E}[\lambda]\,,
\ee
The explicit form of the emissivity function depends on the physical situation being considered. Comparing with the astrophysical model of \cite{Cusin:2017fwz}  we find
\be
\mathcal{E}[\lambda]=\left(\frac{d\tau}{d\lambda}\right) \mathcal{L}_{\Gal}(\vartheta_{\Gal}, \omega_{\Gal})\frac{n_{\Gal}}{\omega^3_{\Gal}}\,, 
\ee
where $\vartheta_{\Gal}$ are a set of parameters effectively describing a given galaxy (mass, metallicity, etc.). 
It follows that eq. (\ref{LL}) can be rewritten as
\be\label{icci}
\frac{df}{d\eta}=\left(\frac{d\tau}{d\eta}\right)\mathcal{L}_{\Gal}(\vartheta_{\Gal}, \omega_{\Gal})\frac{n_{\Gal}}{\omega_{\Gal}^3}\,.
\ee
The relation between proper time and conformal time  \cite{Cusin:2017fwz} is 
\be\label{icc}
\frac{d\tau}{d\eta}=\sqrt{p^{\mu}p^{\nu}g_{\mu\nu}}\frac{d\lambda}{d\eta}=a\left[1+\psi-\bv\cdot \bn\right]\,,
\ee
where $p^{\mu}$ is the spatial projection of the graviton 4-vector. 
%\rut{[NOTE: Here I am quite sure that $p^\mu$ is the 4-velocity of the emitter. Otherwise I don't understand the rhs of this equation...]}

We can now solve eq. (\ref{icci}) replacing  (\ref{icc}) and (\ref{qua}), and then integrating over energy to find an equation for $\Omega_{GW}$
%  and we split as in eq. (\ref{split}). 
The equation for the background becomes
\be
\bar{\Omega}_{GW}(\eta_0)=\int_0^{\eta_0}d\eta\, a(\eta)^5\, \bar{n}_{\Gal}(\eta)\int\, d\eta_{\Gal}\,L_{\Gal}(\vartheta_{\Gal})\,, 
\ee
where we have introduced the integrated luminosity
\be\label{LLG}
L_{\Gal}(\vartheta_{\Gal})=\int d\omega_{\Gal} \mathcal{L}_{\Gal}(\omega_{\Gal}, \vartheta_{\Gal})\,.
\ee
The equation for the perturbed quantity $\delta\Omega_{GW}$ can be written as
\begin{align}
\partial_{\eta}\delta\Omega_{GW}+n^i\partial_i\delta\Omega_{GW}+4\mathcal{H} \delta\Omega_{GW}=&-4\left[\frac{d}{d\eta}(\phi+\bn\cdot \bv)-\dot{\psi}-\dot{\phi}\right]\bar{\Omega}_{GW}\nn\\
&+\bar{n}_{\Gal}a \left(\psi-\bv\cdot \bn+\delta_{\Gal}\right)\int\,d\vartheta_{\Gal}\, L_{\Gal}\,.
\end{align}
This equation can be solved by writing the left hand side as
\be
(l.h.s.)=a^{-4}\frac{d}{d\eta}\left(a^4 \delta\Omega_{GW}\right)\,,
\ee
and after standard manipulations and replacing $\Omega_{GW}\rightarrow 4\pi \Omega_{GW}$, we find
\be
\delta\Omega_{GW}(\bee)=\frac{1}{4\pi\rho_c}\int_0^{\eta_0}d\eta\, a^5\, \bar{n}_{\Gal}(\eta)\left[\delta_{\Gal}+4\phi+\psi-3\bee\cdot \bv+4\int_{\eta}^{\eta_0}d\eta'\, (\dot{\phi}+\dot{\psi})\right]\int d\vartheta_{\Gal}\, L_{\Gal}(\vartheta_{\Gal})\,.
\ee
The corresponding result per logarithmic frequency can be obtained  using eq. (\ref{LLG}), recalling that $\omega_{\Gal}=(1+z_{\Gal}) \omega$ and taking into account redshift perturbations is
\be\label{YY}
\Omega_{GW}(\omega, \bee)=\frac{\omega}{4\pi\rho_c}\int_0^{\eta_0} d\eta\, a^4\bar{n}_{\Gal}(\eta)\left[1+\delta_{\Gal}+4\psi-2\bee\cdot \nabla v+3\int_{\eta}^{\eta_0}d\eta'\, (\dot{\psi}+\dot{\phi})\right]\int d\vartheta_{\Gal} \mathcal{L}_{\Gal}(\omega_{\Gal}, \vartheta_{\Gal})\,, 
\ee
which agrees with the result given in eq. (67) of \cite{Cusin:2017fwz}. 

\subsection{A cosmological background}

For a GW background of cosmological origin, we solve (\ref{LL}) with stochastic initial conditions for the various fields (and no emissivity). One has
\be
\bar{\Omega}_{GW}(\eta_0)=\frac{a(\eta_i)^4}{a(\eta_0)^4} \bar{\Omega}_{GW}(\eta_i)\,,
\ee
and 
\be
\frac{\delta\Omega_{GW}(\eta_0, \bee)}{\bar{\Omega}_{GW}(\eta_0)}=\frac{\delta\Omega_{GW}(\eta_i, \bee)}{\bar{\Omega}_{GW}(\eta_i)}-4\left[\phi_0-\phi_i-\bee\cdot \bv_0+\bee\cdot \bv_i\right]+4\int_{\eta_i}^{\eta_0}d\eta\left(\dot{\phi}+\dot{\psi}\right)\,,
\ee
where $\eta_i$ is the initial time at which the background is produced. The result per units of logarithmic frequency is simply 
\be\label{cosmo final}
\frac{\delta\Omega_{GW}(\eta_0, \bee, \omega)}{\bar{\Omega}_{GW}(\eta_0, \omega)}=\frac{\delta\Omega_{GW}(\eta_i, \bee, \omega)}{\bar{\Omega}_{GW}(\eta_i, \omega)}-4\left[\phi_0-\phi_i-\bee\cdot \bv_0+\bee\cdot \bv_i\right]+4\int_{\eta_i}^{\eta_0}d\eta\left(\dot{\phi}+\dot{\psi}\right)\,. 
\ee

\newpage

\section{The amplitude of fluctuations}\label{new}

\subsection{Astrophysical background}

An order of magnitude estimate of the amplitude of the anisotropies of the astrophysical background with respect to the isotropic component can be obtained by computing
\begin{align}\label{sigma mm}
\sigma_{GW}^2(f)&\equiv \frac{\langle\delta\Omega_{GW}(\bee, f)\delta\Omega_{GW}(\bee, f)\rangle}{\bar{\Omega}^2_{GW}(f)}=\frac{1}{\bar{\Omega}^2_{GW}(f)}\sum_{\ell}\frac{2\ell+1}{2\pi}C_{\ell}(f)\,. 
\end{align}
For the second equality we have used that the correlation function. 
\be\label{corr1}
C(f, \theta)=\langle \delta\Omega_{GW}(f, \bee_1)\delta\Omega_{GW}(f, \bee_2)\rangle\,,
\ee
with $\bee_1\cdot \bee_2=\cos\theta$ can be expanded in  Legendre polynomials as 
\be\label{corr2}
C(f, \theta)\equiv \sum_{\ell} \frac{2\ell+1}{2\pi} C_{\ell}(f) P_{\ell}(\bee_1\cdot \bee_2)\,.
\ee
The angular power spectrum has the following expression, see \cite{Cusin:2017fwz}
\be\label{Cell}
C_{\ell}(f)=\frac{2}{\pi}\int \dd k\,k^2 |\delta\Omega_{\ell}(k, f)|^2\,. 
\ee
Here $k$ is the Fourier variable and $\delta\Omega_{\ell}(k, f)$ is the $\ell$-mode of the the Fourier transform of $\delta\Omega_{\ell}(\bx, f)$. On large scales it is simply given by~\cite{Cusin:2017fwz}  
\be\label{Rkk1}
\delta\Omega_{\ell}(k, f)=\frac{f}{4\pi\rho_c}\int_{0}^{\eta_{\obs}} \dd\eta\,a^5(\eta) \bar{n}_{\Gal}(\eta) b\delta_{k}(\eta) j_{\ell}(k\Delta\eta)\int d\vartheta_{\Gal} \mathcal{L}_{\Gal}(f_{\Gal}, \vartheta_{\Gal})\,, 
\ee
%\rut{[NOTE: it is a bit awkward to have $\theta$ and $\theta_G$ which have completely different meanings. I am tempted to replace the latter by $\vartheta_{\Gal}$]}\\
where $b$ is the (scale-independent) bias and $\delta_k$ the matter over density (in Fourier space). We work under the following hypothesis': 
\begin{enumerate}
\item all galaxies have the same integrated luminosity (i.e. $\mathcal{L}_{\Gal}(f)$ does not depend on $\vartheta_{\Gal}$); 
\item the luminosity (integrated) does not depend on time, i.e. the luminosity per units of frequency scales with a simple redshift factor
 \be\label{Irina}
 \mathcal{L}_{\Gal}(f_{\Gal}, \vartheta_{\Gal})= \mathcal{L}_{\Gal}(f, \vartheta_{\Gal})\frac{df}{df_{\Gal}}=(1+z_{\Gal})^{-1} \mathcal{L}_{\Gal}(f, \vartheta_{\Gal})\,;
 \ee
\item the universe is matter dominated, $a(\eta)=(\eta/\eta_{\obs})^2$;
\item $\delta_m=\delta_{\text{CDM}}$ (we neglect baryons). 
\end{enumerate}
We focus on sub-horizon modes in matter domination, i.e. on $k\gg k_{\text{eq}}$, for which 
%\\ \rut{[NOTE: superhorizon in md is  $k\ll k_{\text{eq}}$ and for these modes the eqn. below is only valid in co-moving gauge.  I think you actually concentrate on SUB horizon modes.]}
\be\label{log}
\delta_{\rm k}(\eta)=\delta_{\rm k}\left(\frac{\eta}{\eta_{\text{eq}}}\right)^2\,,
\ee
with 
\be
\delta_{\rm k}=\frac{9}{10} \Phi_{\rm k}^P\left[-1+6\log(k\eta_{\text{eq}})\right]\,,
\ee
with the primordial power spectrum given by 
\begin{align}\label{primordial}
\Phi^P_{k}=\pi\sqrt{2}\frac{2}{3} k^{-3/2} \frac{A_{\rm S}^{1/2}}{\left[1+\frac{4}{15}R_{\nu}\right]} \left(\frac{k}{k_{\text{ref}}}\right)^{(n_s-1)/2}\,,
 \end{align}
 where $R_\nu$ is the fraction of neutrinos in radiation $R_\nu \equiv \Omega_\nu/(\Omega_\nu+\Omega_\gamma) = \alpha/(1+\alpha)$ 
with $\alpha = N_{\rm eff} 7/8 (4/11)^{4/3}$ and $N_{\rm eff}=3.046$ is the effective number of neutrino species, see e.g. \cite{deSalas:2016ztq}. \footnote{All values for the cosmological parameters are those of  \cite{Ade:2015xua}, explicitly $k_{\text{ref }}= 0.002 \text{Mpc}$, $A_{\rm S}=2.1986\cdot 10^{-9}$, $n_s=0.9652$. }
We use 
\be\label{miab}
\bar{n}_{\Gal}(\eta)=n_{\Gal, \obs}\left(\frac{a_{\obs}}{a}\right)^3=n_{\Gal, \obs}\left(\frac{\eta_{\obs}}{\eta}\right)^6\,.
\ee
and we assume the bias to be scale-independent and to scale as $\propto \sqrt{1+z}$  \cite{Marin:2013bbb, Rassat:2008ja}:
\be\label{miaf}
b\sim b_{\obs}\sqrt{1+z}=b_{\obs}\left(\frac{\eta_{\obs}}{\eta}\right)\,,
\ee
with $b_{\obs}\sim 1$. 
Putting all these ingredients together and making use of the following asymptotic behavior of the spherical Bessel function
\be\label{as}
j_{\ell}(x) \sim \sqrt{\frac{\pi}{2\ell+1}}\delta\left(\ell+\frac{1}{2} -x\right)+\mathcal{O}(1/\ell^2)\,,
\ee
eq. (\ref{Rkk1}) simplifies to 
\be\label{cuculo}
\delta\hat{\Omega}_{\ell}(k, f)=\frac{f}{4\pi \rho_c} \mathcal{L}_{\Gal}(f) b_{\obs} n_{\Gal, \obs} \frac{1}{k}\left( \frac{\eta_{\obs}}{\eta_{\text{eq}}}\right)^2\left( \frac{\bar{\eta}}{\eta_{\obs}}\right)^7\sqrt{\frac{\pi}{2\ell+1}} \delta_k\,,
\ee
with
\be\label{etabar}
\bar{\eta}=\eta_{\obs}-\left(\ell+\frac{1}{2}\right) \frac{1}{k}\,,
\ee
which can be approximated to $\bar{\eta}=\eta_{\obs}$ for large enough angular scales.  Then 
%\be
%\delta\hat{\Omega}_{\ell}(k, f)=\frac{f}{4\pi \rho_c} \mathcal{L}_{\Gal}(f) b_{\obs} n_{\Gal, \obs} \frac{1}{k}\left( \frac{\eta_{\obs}}{\eta_{\text{eq}}}\right)^2\sqrt{\frac{\pi}{2\ell+1}} \delta_k\,,
%\ee
%and 
\be\label{Clest1}
C_{\ell}(f)=\left(\ell+\frac{1}{2}\right)^{-1} \left[\frac{f}{4\pi \rho_c} \mathcal{L}_{\Gal}(f) b_{\obs} n_{\Gal, \obs}\right]^2\left(\frac{\eta_{\obs}}{\eta_{\rm eq}}\right)^4\int_{\rm k_{eq}} d{\rm k}\, |\delta_{\rm k}|^2\,,
\ee
with\footnote{We neglect the dependence on the logarithm in eq. (\ref{log}).}
\be\label{Clest2}
\int_{\rm k_{eq}} d{\rm k}\, |\delta_{\rm k}|^2\simeq \pi^2 \frac{8}{9} \left(\frac{9}{10}\right)^2\frac{1}{3-n_s} \left[\frac{A^{1/2}_{\rm S}}{1+\frac{4}{15}R_{\nu}}\right]^2 \left(\frac{k_{\rm ref}}{k_{\rm eq}}\right)^{1- n_s}\left(\frac{\eta_{\rm eq}}{\eta_{\obs}}\right)^2 \eta_{\obs}^2\,.
\ee
While for the isotropic component we have
\be\label{back}
\bar{\Omega}^2_{GW}(f)\simeq \frac{1}{25} \left[\frac{f}{4\pi \rho_c} \mathcal{L}_{\Gal}(f) n_{\Gal, \obs}\right]^2\eta_{\obs}^2\,,
\ee
Going back to eq. (\ref{sigma}) and using eqs. (\ref{Clest1}), (\ref{Clest2}) and (\ref{back}), we find
\be\label{final sigma}
\sigma^2_{GW}\simeq 18\pi\,b_{\obs}^2 \left[\frac{A^{1/2}_{\rm S}}{1+\frac{4}{15}R_{\nu}}\right]^2\left(\frac{k_{\rm ref}}{k_{\rm eq}}\right)^{1- n_s}\left(\frac{\eta_{\obs}}{\eta_{\rm eq}}\right)^2\sim 10^{-4}\,.
\ee
We can conclude that the relative amplitude of fluctuations with respect to the isotropic component, for a generic astrophysical background is of order
\be\label{final sigma 2}
\frac{\delta \Omega_{GW}}{\bar{\Omega}_{GW}}\sim \sigma_{GW} \approx 10^{-2}\,.
\ee
This result applies to any component of the astrophysical background as long as the luminosity function $\mathcal{L}_{\Gal}$ depends on time only through the redshifted frequency (see eq. \ref{Irina}), which is the case e.g. for a background of mergers in the LIGO band. The estimate can be refined by assuming for each contribution to the background (black hole mergers, supernovae, \dots) a specific frequency dependence of the luminosity function. Eq. \ref{final sigma 2} is in agreement with the numerical results obtained in \cite{Cusin:2018rsq} for the case of a background of black hole mergers. 

Note that  (\ref{final sigma 2}) is an estimate of the amount of anisotropy today, at $\eta=\eta_0$.  The analogous quantity at a different time $\eta_1<\eta_0$, can be worked out in a similar way: in the definition of the energy density (\ref{YY}), the integral over time runs up to $\eta_1$ and eq. (\ref{cuculo})  has to be evaluated at $\bar{\eta}_1=\eta_1-(\ell+1/2)/k$. For $\eta_1\gg \eta_{\text{eq}}$, one has $\eta_1/\eta_0\gg  (\ell+1/2)/(\eta_0 k_{\text{eq}})\gg (\ell+1/2)/(\eta_0 k)$, hence $\bar{\eta_1}/\eta_0\simeq \eta_1/\eta_0$ and the final result for the fluctuations is suppressed by a factor $(\eta_1/\eta_0)^7$ with respect to (\ref{final sigma 2}). In section \ref{visibility 1} we consider a simplified framework where most of astrophysical sources are located at redshift $z=1.5$. For this situation $\delta\Omega_{GW}/\bar{\Omega}_{GW}(\eta_1)\simeq(\eta_1/\eta_0)^7\delta\Omega_{GW}/\bar{\Omega}_{GW}(\eta_0)\simeq(a_0/a_1)^{7/2}\delta\Omega_{GW}/\bar{\Omega}_{GW}(\eta_0)\simeq 10^{-3}$. 

\subsection{Cosmological background}
%{\color{red}THESE TWO SUBSECTIONS SEEM TO BE IN THE INVERSE ORDER TO THE ONES IN THE PREVIOUS APPENDIX. THIS LOOK ODD}

The anisotropies of a cosmological GW background are a tracer of the temperature anisotropies of the CMB. We therefore expect $\delta\Omega_{GW}/\bar{\Omega}_{GW}(\eta_0)\simeq 10^{-5}$. If we consider  $\eta_1<\eta_0$, the result changes slightly  due to the change of the integration bounds of the integrated Sachs Wolfe term in eq. (\ref{cosmo final}). As a first approximation, we neglect this correction, assuming that the value $10^{-5}$ stays the same for any redshift.

\newpage

 \bibliographystyle{utphys}
\bibliography{myrefs}

\providecommand{\href}[2]{#2}\begingroup\raggedright\begin{thebibliography}{100}

\bibitem{Penzias:1965wn}
A.~A. Penzias and R.~W. Wilson, ``{A Measurement of excess antenna temperature
  at 4080-Mc/s},'' {\em Astrophys. J.} {\bf 142} (1965) 419--421.

\bibitem{2001ARA&A..39..249H}
M.~G. {Hauser} and E.~{Dwek}, ``{The Cosmic Infrared Background: Measurements
  and Implications},'' {\em Annual Review of Astronomy and Astrophysics} {\bf
  39} (2001) 249--307, \href{http://xxx.lanl.gov/abs/astro-ph/0105539}{{\tt
  astro-ph/0105539}}.

\bibitem{1967ApJ...148..377P}
R.~B. {Partridge} and P.~J.~E. {Peebles}, ``{Are Young Galaxies Visible? II.
  The Integrated Background},'' {\em Astrophys. J.} {\bf 148} (May, 1967) 377.

\bibitem{1991Natur.353..315S}
T.~{Shanks}, I.~{Georgantopoulos}, G.~C. {Stewart}, K.~A. {Pounds}, B.~J.
  {Boyle}, and R.~E. {Griffiths}, ``{The origin of the cosmic X-ray
  background},'' {\em Nature} {\bf 353} (Sept., 1991) 315--320.

\bibitem{2006ARNPS..56..137H}
S.~{Hannestad}, ``{Primordial Neutrinos},'' {\em Annual Review of Nuclear and
  Particle Science} {\bf 56} (Nov., 2006) 137--161,
  \href{http://xxx.lanl.gov/abs/hep-ph/0602058}{{\tt hep-ph/0602058}}.

\bibitem{PeterUzan2005}
P.~Peter and J.-P. Uzan, {\em {Primordial Cosmology}}.
\newblock Oxford Graduate Texts. Oxford University Press, 2005.

\bibitem{Dufaux:2007pt}
J.~F. Dufaux, A.~Bergman, G.~N. Felder, L.~Kofman, and J.-P. Uzan, ``{Theory
  and Numerics of Gravitational Waves from Preheating after Inflation},'' {\em
  Phys. Rev.} {\bf D76} (2007) 123517,
  \href{http://xxx.lanl.gov/abs/0707.0875}{{\tt 0707.0875}}.

\bibitem{Vilenkin:1981bx}
A.~Vilenkin, ``{Gravitational radiation from cosmic strings},'' {\em Phys.
  Lett.} {\bf B107} (1981) 47--50.

\bibitem{Hogan:1984is}
C.~J. Hogan and M.~J. Rees, ``{Gravitational interactions of cosmic strings},''
  {\em Nature} {\bf 311} (1984) 109--113.

\bibitem{Vachaspati:1984gt}
T.~Vachaspati and A.~Vilenkin, ``{Gravitational Radiation from Cosmic
  Strings},'' {\em Phys. Rev.} {\bf D31} (1985) 3052.

\bibitem{Caldwell:1991jj}
R.~R. Caldwell and B.~Allen, ``{Cosmological constraints on cosmic string
  gravitational radiation},'' {\em Phys. Rev.} {\bf D45} (1992) 3447--3468.

\bibitem{Kuroyanagi:2016ugi}
S.~Kuroyanagi, K.~Takahashi, N.~Yonemaru, and H.~Kumamoto, ``{Anisotropies in
  the gravitational wave background as a probe of the cosmic string network},''
  {\em Phys. Rev.} {\bf D95} (2017), no.~4 043531,
  \href{http://xxx.lanl.gov/abs/1604.00332}{{\tt 1604.00332}}.

\bibitem{Caprini:2009fx}
C.~Caprini, R.~Durrer, T.~Konstandin, and G.~Servant, ``{General Properties of
  the Gravitational Wave Spectrum from Phase Transitions},'' {\em Phys. Rev.}
  {\bf D79} (2009) 083519, \href{http://xxx.lanl.gov/abs/0901.1661}{{\tt
  0901.1661}}.

\bibitem{Caprini:2015zlo}
C.~Caprini {\em et.~al.}, ``{Science with the space-based interferometer eLISA.
  II: Gravitational waves from cosmological phase transitions},'' {\em JCAP}
  {\bf 1604} (2016), no.~04 001, \href{http://xxx.lanl.gov/abs/1512.06239}{{\tt
  1512.06239}}.

\bibitem{Caprini:2001nb}
C.~Caprini and R.~Durrer, ``{Gravitational wave production: A Strong constraint
  on primordial magnetic fields},'' {\em Phys. Rev.} {\bf D65} (2001) 023517,
  \href{http://xxx.lanl.gov/abs/astro-ph/0106244}{{\tt astro-ph/0106244}}.

\bibitem{Binetruy:2012ze}
P.~Binetruy, A.~Bohe, C.~Caprini, and J.-F. Dufaux, ``{Cosmological Backgrounds
  of Gravitational Waves and eLISA/NGO: Phase Transitions, Cosmic Strings and
  Other Sources},'' {\em JCAP} {\bf 1206} (2012) 027,
  \href{http://xxx.lanl.gov/abs/1201.0983}{{\tt 1201.0983}}.

\bibitem{Caprini:2018mtu}
C.~Caprini and D.~G. Figueroa, ``{Cosmological Backgrounds of Gravitational
  Waves},'' \href{http://xxx.lanl.gov/abs/1801.04268}{{\tt 1801.04268}}.

\bibitem{Maggiore:1999vm}
M.~Maggiore, ``{Gravitational wave experiments and early universe cosmology},''
  {\em Phys. Rept.} {\bf 331} (2000) 283--367,
  \href{http://xxx.lanl.gov/abs/gr-qc/9909001}{{\tt gr-qc/9909001}}.

\bibitem{Buonanno:2014aza}
A.~Buonanno and B.~S. Sathyaprakash, ``{Sources of Gravitational Waves: Theory
  and Observations},'' \href{http://xxx.lanl.gov/abs/1410.7832}{{\tt
  1410.7832}}.

\bibitem{TheLIGOScientific:2016wyq}
{\bf Virgo, LIGO Scientific} Collaboration, B.~P. Abbott {\em et.~al.},
  ``{GW150914: Implications for the stochastic gravitational wave background
  from binary black holes},'' {\em Phys. Rev. Lett.} {\bf 116} (2016), no.~13
  131102, \href{http://xxx.lanl.gov/abs/1602.03847}{{\tt 1602.03847}}.

\bibitem{Regimbau:2016ike}
T.~Regimbau, M.~Evans, N.~Christensen, E.~Katsavounidis, B.~Sathyaprakash, and
  S.~Vitale, ``{Digging deeper: Observing primordial gravitational waves below
  the binary black hole produced stochastic background},''
  \href{http://xxx.lanl.gov/abs/1611.08943}{{\tt 1611.08943}}.

\bibitem{Mandic:2016lcn}
V.~Mandic, S.~Bird, and I.~Cholis, ``{Stochastic Gravitational-Wave Background
  due to Primordial Binary Black Hole Mergers},'' {\em Phys. Rev. Lett.} {\bf
  117} (2016), no.~20 201102, \href{http://xxx.lanl.gov/abs/1608.06699}{{\tt
  1608.06699}}.

\bibitem{Dvorkin:2016okx}
I.~Dvorkin, J.-P. Uzan, E.~Vangioni, and J.~Silk, ``{Synthetic model of the
  gravitational wave background from evolving binary compact objects},'' {\em
  Phys. Rev.} {\bf D94} (2016), no.~10 103011,
  \href{http://xxx.lanl.gov/abs/1607.06818}{{\tt 1607.06818}}.

\bibitem{Nakazato:2016nkj}
K.~Nakazato, Y.~Niino, and N.~Sago, ``{Gravitational-Wave Background from
  Binary Mergers and Metallicity Evolution of Galaxies},'' {\em Astrophys. J.}
  {\bf 832} (2016), no.~2 146, \href{http://xxx.lanl.gov/abs/1605.02146}{{\tt
  1605.02146}}.

\bibitem{Dvorkin:2016wac}
I.~Dvorkin, E.~Vangioni, J.~Silk, J.-P. Uzan, and K.~A. Olive,
  ``{Metallicity-constrained merger rates of binary black holes and the
  stochastic gravitational wave background},'' {\em Mon. Not. Roy. Astron.
  Soc.} {\bf 461} (2016), no.~4 3877--3885,
  \href{http://xxx.lanl.gov/abs/1604.04288}{{\tt 1604.04288}}.

\bibitem{Evangelista:2014oba}
E.~F.~D. Evangelista and J.~C.~N. Araujo, ``{The Gravitational Wave Background
  from Coalescing Compact Binaries: A New Method},'' {\em Braz. J. Phys.} {\bf
  44} (2014), no.~6 824--831, \href{http://xxx.lanl.gov/abs/1504.06605}{{\tt
  1504.06605}}.

\bibitem{Kelley:2017lek}
L.~Z. Kelley, L.~Blecha, L.~Hernquist, and A.~Sesana, ``{The Gravitational Wave
  Background from Massive Black Hole Binaries in Illustris: spectral features
  and time to detection with pulsar timing arrays},''
  \href{http://xxx.lanl.gov/abs/1702.02180}{{\tt 1702.02180}}.

\bibitem{Surace:2015ppq}
M.~Surace, K.~D. Kokkotas, and P.~Pnigouras, ``{The stochastic background of
  gravitational waves due to the $f$-mode instability in neutron stars},'' {\em
  Astron. Astrophys.} {\bf 586} (2016) A86,
  \href{http://xxx.lanl.gov/abs/1512.02502}{{\tt 1512.02502}}.

\bibitem{Talukder:2014eba}
D.~Talukder, E.~Thrane, S.~Bose, and T.~Regimbau, ``{Measuring neutron-star
  ellipticity with measurements of the stochastic gravitational-wave
  background},'' {\em Phys. Rev.} {\bf D89} (2014), no.~12 123008,
  \href{http://xxx.lanl.gov/abs/1404.4025}{{\tt 1404.4025}}.

\bibitem{Lasky:2013jfa}
P.~D. Lasky, M.~F. Bennett, and A.~Melatos, ``{Stochastic gravitational wave
  background from hydrodynamic turbulence in differentially rotating neutron
  stars},'' {\em Phys. Rev.} {\bf D87} (2013), no.~6 063004,
  \href{http://xxx.lanl.gov/abs/1302.6033}{{\tt 1302.6033}}.

\bibitem{Crocker:2017agi}
K.~Crocker, T.~Prestegard, V.~Mandic, T.~Regimbau, K.~Olive, and E.~Vangioni,
  ``{A Systematic Study of the Stochastic Gravitational-Wave Background due to
  Stellar Core Collapse},'' \href{http://xxx.lanl.gov/abs/1701.02638}{{\tt
  1701.02638}}.

\bibitem{Crocker:2015taa}
K.~Crocker, V.~Mandic, T.~Regimbau, K.~Belczynski, W.~Gladysz, K.~Olive,
  T.~Prestegard, and E.~Vangioni, ``{Model of the stochastic gravitational-wave
  background due to core collapse to black holes},'' {\em Phys. Rev.} {\bf D92}
  (2015), no.~6 063005, \href{http://xxx.lanl.gov/abs/1506.02631}{{\tt
  1506.02631}}.

\bibitem{Kowalska:2012ba}
I.~Kowalska, T.~Bulik, and K.~Belczynski, ``{Gravitational wave background from
  population III binaries},'' {\em Astron. Astrophys.} {\bf 541} (2012) A120,
  \href{http://xxx.lanl.gov/abs/1202.3346}{{\tt 1202.3346}}.

\bibitem{Abbott:2016blz}
{\bf Virgo, LIGO Scientific} Collaboration, B.~P. Abbott {\em et.~al.},
  ``{Observation of Gravitational Waves from a Binary Black Hole Merger},''
  {\em Phys. Rev. Lett.} {\bf 116} (2016), no.~6 061102,
  \href{http://xxx.lanl.gov/abs/1602.03837}{{\tt 1602.03837}}.

\bibitem{Abbott:2016nmj}
{\bf Virgo, LIGO Scientific} Collaboration, B.~P. Abbott {\em et.~al.},
  ``{GW151226: Observation of Gravitational Waves from a 22-Solar-Mass Binary
  Black Hole Coalescence},'' {\em Phys. Rev. Lett.} {\bf 116} (2016), no.~24
  241103, \href{http://xxx.lanl.gov/abs/1606.04855}{{\tt 1606.04855}}.

\bibitem{TheLIGOScientific:2016pea}
{\bf Virgo, LIGO Scientific} Collaboration, B.~P. Abbott {\em et.~al.},
  ``{Binary Black Hole Mergers in the first Advanced LIGO Observing Run},''
  {\em Phys. Rev.} {\bf X6} (2016), no.~4 041015,
  \href{http://xxx.lanl.gov/abs/1606.04856}{{\tt 1606.04856}}.

\bibitem{Abbott:2017vtc}
{\bf VIRGO, LIGO Scientific} Collaboration, B.~P. Abbott {\em et.~al.},
  ``{GW170104: Observation of a 50-Solar-Mass Binary Black Hole Coalescence at
  Redshift 0.2},'' {\em Phys. Rev. Lett.} {\bf 118} (2017), no.~22 221101,
  \href{http://xxx.lanl.gov/abs/1706.01812}{{\tt 1706.01812}}.

\bibitem{Abbott:2017gyy}
{\bf Virgo, LIGO Scientific} Collaboration, B.~P. Abbott {\em et.~al.},
  ``{GW170608: Observation of a 19-solar-mass Binary Black Hole Coalescence},''
  {\em Astrophys. J.} {\bf 851} (2017), no.~2 L35,
  \href{http://xxx.lanl.gov/abs/1711.05578}{{\tt 1711.05578}}.

\bibitem{Abbott:2017oio}
{\bf Virgo, LIGO Scientific} Collaboration, B.~P. Abbott {\em et.~al.},
  ``{GW170814: A three-detector observation of gravitational waves from a
  binary black hole coalescence},'' {\em Submitted to: Phys. Rev. Lett.} (2017)
  \href{http://xxx.lanl.gov/abs/1709.09660}{{\tt 1709.09660}}.

\bibitem{TheLIGOScientific:2017qsa}
{\bf Virgo, LIGO Scientific} Collaboration, B.~Abbott {\em et.~al.},
  ``{GW170817: Observation of Gravitational Waves from a Binary Neutron Star
  Inspiral},'' {\em Phys. Rev. Lett.} {\bf 119} (2017), no.~16 161101,
  \href{http://xxx.lanl.gov/abs/1710.05832}{{\tt 1710.05832}}.

\bibitem{Abbott:2017xzg}
{\bf Virgo, LIGO Scientific} Collaboration, B.~P. Abbott {\em et.~al.},
  ``{GW170817: Implications for the Stochastic Gravitational-Wave Background
  from Compact Binary Coalescences},'' {\em Phys. Rev. Lett.} {\bf 120} (2018),
  no.~9 091101, \href{http://xxx.lanl.gov/abs/1710.05837}{{\tt 1710.05837}}.

\bibitem{Allen:1996vm}
B.~Allen, ``{The Stochastic gravity wave background: Sources and detection},''
  in {\em {Relativistic gravitation and gravitational radiation. Proceedings,
  School of Physics, Les Houches, France, September 26-October 6, 1995}},
  pp.~373--417, 1996.
\newblock \href{http://xxx.lanl.gov/abs/gr-qc/9604033}{{\tt gr-qc/9604033}}.

\bibitem{Smith:2006nka}
T.~L. Smith, E.~Pierpaoli, and M.~Kamionkowski, ``{A new cosmic microwave
  background constraint to primordial gravitational waves},'' {\em Phys. Rev.
  Lett.} {\bf 97} (2006) 021301,
  \href{http://xxx.lanl.gov/abs/astro-ph/0603144}{{\tt astro-ph/0603144}}.

\bibitem{Henrot-Versille:2014jua}
S.~Henrot-Versille {\em et.~al.}, ``{Improved constraint on the primordial
  gravitational-wave density using recent cosmological data and its impact on
  cosmic string models},'' {\em Class. Quant. Grav.} {\bf 32} (2015), no.~4
  045003, \href{http://xxx.lanl.gov/abs/1408.5299}{{\tt 1408.5299}}.

\bibitem{Shannon:2013wma}
R.~M. Shannon {\em et.~al.}, ``{Gravitational-wave Limits from Pulsar Timing
  Constrain Supermassive Black Hole Evolution},'' {\em Science} {\bf 342}
  (2013), no.~6156 334--337, \href{http://xxx.lanl.gov/abs/1310.4569}{{\tt
  1310.4569}}.

\bibitem{Allen:1996gp}
B.~Allen and A.~C. Ottewill, ``{Detection of anisotropies in the gravitational
  wave stochastic background},'' {\em Phys. Rev.} {\bf D56} (1997) 545--563,
  \href{http://xxx.lanl.gov/abs/gr-qc/9607068}{{\tt gr-qc/9607068}}.

\bibitem{Cornish:2001hg}
N.~J. Cornish, ``{Mapping the gravitational wave background},'' {\em Class.
  Quant. Grav.} {\bf 18} (2001) 4277--4292,
  \href{http://xxx.lanl.gov/abs/astro-ph/0105374}{{\tt astro-ph/0105374}}.

\bibitem{Mitra:2007mc}
S.~Mitra, S.~Dhurandhar, T.~Souradeep, A.~Lazzarini, V.~Mandic, S.~Bose, and
  S.~Ballmer, ``{Gravitational wave radiometry: Mapping a stochastic
  gravitational wave background},'' {\em Phys. Rev.} {\bf D77} (2008) 042002,
  \href{http://xxx.lanl.gov/abs/0708.2728}{{\tt 0708.2728}}.

\bibitem{Thrane:2009fp}
E.~Thrane, S.~Ballmer, J.~D. Romano, S.~Mitra, D.~Talukder, S.~Bose, and
  V.~Mandic, ``{Probing the anisotropies of a stochastic gravitational-wave
  background using a network of ground-based laser interferometers},'' {\em
  Phys. Rev.} {\bf D80} (2009) 122002,
  \href{http://xxx.lanl.gov/abs/0910.0858}{{\tt 0910.0858}}.

\bibitem{Romano:2015uma}
J.~D. Romano, S.~R. Taylor, N.~J. Cornish, J.~Gair, C.~M.~F. Mingarelli, and
  R.~van Haasteren, ``{Phase-coherent mapping of gravitational-wave backgrounds
  using ground-based laser interferometers},'' {\em Phys. Rev.} {\bf D92}
  (2015), no.~4 042003, \href{http://xxx.lanl.gov/abs/1505.07179}{{\tt
  1505.07179}}.

\bibitem{Romano:2016dpx}
J.~D. Romano and N.~J. Cornish, ``{Detection methods for stochastic
  gravitational-wave backgrounds: A unified treatment},''
  \href{http://xxx.lanl.gov/abs/1608.06889}{{\tt 1608.06889}}.

\bibitem{TheLIGOScientific:2016xzw}
{\bf Virgo, LIGO Scientific} Collaboration, B.~P. Abbott {\em et.~al.},
  ``{Directional Limits on Persistent Gravitational Waves from Advanced LIGO?s
  First Observing Run},'' {\em Phys. Rev. Lett.} {\bf 118} (2017), no.~12
  121102, \href{http://xxx.lanl.gov/abs/1612.02030}{{\tt 1612.02030}}.

\bibitem{Mingarelli:2013dsa}
C.~M.~F. Mingarelli, T.~Sidery, I.~Mandel, and A.~Vecchio, ``{Characterizing
  gravitational wave stochastic background anisotropy with pulsar timing
  arrays},'' {\em Phys. Rev.} {\bf D88} (2013), no.~6 062005,
  \href{http://xxx.lanl.gov/abs/1306.5394}{{\tt 1306.5394}}.

\bibitem{Taylor:2013esa}
S.~R. Taylor and J.~R. Gair, ``{Searching For Anisotropic Gravitational-wave
  Backgrounds Using Pulsar Timing Arrays},'' {\em Phys. Rev.} {\bf D88} (2013)
  084001, \href{http://xxx.lanl.gov/abs/1306.5395}{{\tt 1306.5395}}.

\bibitem{Gair:2014rwa}
J.~Gair, J.~D. Romano, S.~Taylor, and C.~M.~F. Mingarelli, ``{Mapping
  gravitational-wave backgrounds using methods from CMB analysis: Application
  to pulsar timing arrays},'' {\em Phys. Rev.} {\bf D90} (2014), no.~8 082001,
  \href{http://xxx.lanl.gov/abs/1406.4664}{{\tt 1406.4664}}.

\bibitem{Moore:2014lga}
C.~J. Moore, R.~H. Cole, and C.~P.~L. Berry, ``{Gravitational-wave sensitivity
  curves},'' {\em Class. Quant. Grav.} {\bf 32} (2015), no.~1 015014,
  \href{http://xxx.lanl.gov/abs/1408.0740}{{\tt 1408.0740}}.

\bibitem{Evans:2016mbw}
{\bf LIGO Scientific} Collaboration, B.~P. Abbott {\em et.~al.}, ``{Exploring
  the Sensitivity of Next Generation Gravitational Wave Detectors},'' {\em
  Class. Quant. Grav.} {\bf 34} (2017), no.~4 044001,
  \href{http://xxx.lanl.gov/abs/1607.08697}{{\tt 1607.08697}}.

\bibitem{Regimbau:2011rp}
T.~Regimbau, ``{The astrophysical gravitational wave stochastic background},''
  {\em Res. Astron. Astrophys.} {\bf 11} (2011) 369--390,
  \href{http://xxx.lanl.gov/abs/1101.2762}{{\tt 1101.2762}}.

\bibitem{Cusin:2017fwz}
G.~Cusin, C.~Pitrou, and J.-P. Uzan, ``{Anisotropy of the astrophysical
  gravitational wave background: Analytic expression of the angular power
  spectrum and correlation with cosmological observations},'' {\em Phys. Rev.}
  {\bf D96} (2017), no.~10 103019,
  \href{http://xxx.lanl.gov/abs/1704.06184}{{\tt 1704.06184}}.

\bibitem{Cusin:2017mjm}
G.~Cusin, C.~Pitrou, and J.-P. Uzan, ``{The signal of the gravitational wave
  background and the angular correlation of its energy density},'' {\em Phys.
  Rev.} {\bf D97} (2018), no.~12 123527,
  \href{http://xxx.lanl.gov/abs/1711.11345}{{\tt 1711.11345}}.

\bibitem{Cusin:2018rsq}
G.~Cusin, I.~Dvorkin, C.~Pitrou, and J.-P. Uzan, ``{First predictions of the
  angular power spectrum of the astrophysical gravitational wave background},''
  {\em Phys. Rev. Lett.} {\bf 120} (2018) 231101,
  \href{http://xxx.lanl.gov/abs/1803.03236}{{\tt 1803.03236}}.

\bibitem{Geller:2018mwu}
M.~Geller, A.~Hook, R.~Sundrum, and Y.~Tsai, ``{Primordial Anisotropies in the
  Gravitational Wave Background from Cosmological Phase Transitions},''
  \href{http://xxx.lanl.gov/abs/1803.10780}{{\tt 1803.10780}}.

\bibitem{Taylor2015}
S.~R. Taylor {\em et.~al.}, ``{Limits on anisotropy in the nanohertz stochastic
  gravitational-wave background},'' {\em Phys. Rev. Lett.} {\bf 115} (2015),
  no.~4 041101, \href{http://xxx.lanl.gov/abs/1506.08817}{{\tt 1506.08817}}.

\bibitem{Sesana2008}
A.~Sesana, A.~Vecchio, and C.~N. Colacino, ``{The stochastic gravitational-wave
  background from massive black hole binary systems: implications for
  observations with Pulsar Timing Arrays},'' {\em Mon. Not. Roy. Astron. Soc.}
  {\bf 390} (2008) 192, \href{http://xxx.lanl.gov/abs/0804.4476}{{\tt
  0804.4476}}.

\bibitem{TheLIGOScientific:2016dpb}
{\bf Virgo, LIGO Scientific} Collaboration, B.~P. Abbott {\em et.~al.},
  ``{Upper Limits on the Stochastic Gravitational-Wave Background from Advanced
  LIGO First Observing Run},'' {\em Phys. Rev. Lett.} {\bf 118} (2017), no.~12
  121101, \href{http://xxx.lanl.gov/abs/1612.02029}{{\tt 1612.02029}}.
  [Erratum: Phys. Rev. Lett.119,no.2,029901(2017)].

\bibitem{Bonvin:2011bg}
C.~Bonvin and R.~Durrer, ``{What galaxy surveys really measure},'' {\em Phys.
  Rev.} {\bf D84} (2011) 063505, \href{http://xxx.lanl.gov/abs/1105.5280}{{\tt
  1105.5280}}.

\bibitem{Maggiore:2018sht}
M.~Maggiore, {\em {Gravitational Waves. Vol. 2: Astrophysics and Cosmology}}.
\newblock Oxford University Press, 2018.

\bibitem{BHmonography}
J.~A.~H. {Futterman}, F.~A. {Handler}, and R.~A. {Matzner}, {\em {Scattering
  from Black Holes}}.
\newblock Cambridge University Press, June, 2009.

\bibitem{Matzner:1977dn}
R.~A. Matzner and M.~P. Ryan, ``{Low Frequency Limit of Gravitational
  Scattering},'' {\em Phys. Rev.} {\bf D16} (1977) 1636--1642.

\bibitem{Westervelt:1971pm}
P.~J. Westervelt, ``{Scattering of electromagnetic and gravitational waves by a
  static gravitational field - comparison between the classical
  (general-relativistic) and quantum field-theoretic results},'' {\em Phys.
  Rev.} {\bf D3} (1971) 2319--2324.

\bibitem{Peters:1976jx}
P.~C. Peters, ``{Differential Cross-Sections for Weak Field Gravitational
  Scattering},'' {\em Phys. Rev.} {\bf D13} (1976) 775--777.

\bibitem{Sanchez:1976fcl}
N.~G. Sanchez, ``{Scattering of scalar waves from a Schwarzschild black
  hole},'' {\em J. Math. Phys.} {\bf 17} (1976), no.~5 688.

\bibitem{DeLogi:1977dp}
W.~K. De~Logi and S.~J. Kovacs, ``{Gravitational Scattering of Zero Rest Mass
  Plane Waves},'' {\em Phys. Rev.} {\bf D16} (1977) 237--244.

\bibitem{Doran:2001ag}
C.~Doran and A.~Lasenby, ``{Perturbation theory calculation of the black hole
  elastic scattering cross-section},'' {\em Phys. Rev.} {\bf D66} (2002)
  024006, \href{http://xxx.lanl.gov/abs/gr-qc/0106039}{{\tt gr-qc/0106039}}.

\bibitem{Holstein:2006bh}
B.~R. Holstein, ``{Graviton Physics},'' {\em Am. J. Phys.} {\bf 74} (2006)
  1002--1011, \href{http://xxx.lanl.gov/abs/gr-qc/0607045}{{\tt
  gr-qc/0607045}}.

\bibitem{Dolan:2007ut}
S.~R. Dolan, ``{Scattering of long-wavelength gravitational waves},'' {\em
  Phys. Rev.} {\bf D77} (2008) 044004,
  \href{http://xxx.lanl.gov/abs/0710.4252}{{\tt 0710.4252}}.

\bibitem{Guadagnini:2008ha}
E.~Guadagnini, ``{Gravitons scattering from classical matter},'' {\em Class.
  Quant. Grav.} {\bf 25} (2008) 095012,
  \href{http://xxx.lanl.gov/abs/0803.2855}{{\tt 0803.2855}}.

\bibitem{Seto}
N.~Seto and A.~Taruya, ``Polarization analysis of gravitational-wave
  backgrounds from the correlation signals of ground-based interferometers:
  Measuring a circular-polarization mode,'' {\em Phys. Rev. D} {\bf 77} (May,
  2008) 103001.

\bibitem{Gubitosi:2016yoq}
G.~Gubitosi and J.~Magueijo, ``{Correlation between opposite-helicity
  gravitons: Imprints on gravity-wave and microwave backgrounds},'' {\em Phys.
  Rev.} {\bf D95} (2017), no.~2 023520,
  \href{http://xxx.lanl.gov/abs/1610.05702}{{\tt 1610.05702}}.

\bibitem{Kato:2015bye}
R.~Kato and J.~Soda, ``{Probing circular polarization in stochastic
  gravitational wave background with pulsar timing arrays},'' {\em Phys. Rev.}
  {\bf D93} (2016), no.~6 062003,
  \href{http://xxx.lanl.gov/abs/1512.09139}{{\tt 1512.09139}}.

\bibitem{ruthBook}
R.~Durrer, {\em The Cosmic Microwave Background}.
\newblock Cambridge University Press, 2008.

\bibitem{Maggiore:1900zz}
M.~Maggiore, {\em {Gravitational Waves. Vol. 1: Theory and Experiments}}.
\newblock Oxford Master Series in Physics. Oxford University Press, 2007.

\bibitem{2018arXiv180703821S}
F.~R.~N. {Schneider}, O.~H. {Ram{\'{\i}}rez-Agudelo}, F.~{Tramper}, J.~M.
  {Bestenlehner}, N.~{Castro}, H.~{Sana}, C.~J. {Evans},
  C.~{Sab{\'{\i}}n-Sanjuli{\'a}n}, S.~{Sim{\'o}n-D{\'{\i}}az}, N.~{Langer},
  L.~{Fossati}, G.~{Gr{\"a}fener}, P.~A. {Crowther}, S.~E. {de Mink}, A.~{de
  Koter}, M.~{Gieles}, A.~{Herrero}, R.~G. {Izzard}, V.~{Kalari}, R.~S.
  {Klessen}, D.~J. {Lennon}, L.~{Mahy}, J.~{Ma{\'{\i}}z Apell{\'a}niz},
  N.~{Markova}, J.~T. {van Loon}, J.~S. {Vink}, and N.~R. {Walborn}, ``{The
  VLT-FLAMES Tarantula Survey. XXIX. Massive star formation in the local 30
  Doradus starburst},'' {\em ArXiv e-prints} (July, 2018)
  \href{http://xxx.lanl.gov/abs/1807.03821}{{\tt 1807.03821}}.

\bibitem{Zombeck-book}
M.~V. Zombeck, {\em {Handbook of Space Astronomy and Astrophysics (2nd ed.)}}.
\newblock Cambridge University Press, 1990.

\bibitem{2010A&ARv..18...67T}
G.~{Torres}, J.~{Andersen}, and A.~{Gim{\'e}nez}, ``{Accurate masses and radii
  of normal stars: modern results and applications},'' {\em The Astronomy and
  Astrophysics Review} {\bf 18} (Feb., 2010) 67--126,
  \href{http://xxx.lanl.gov/abs/0908.2624}{{\tt 0908.2624}}.

\bibitem{1935MNRAS..95..207C}
S.~{Chandrasekhar}, ``{The highly collapsed configurations of a stellar mass
  (Second paper)},'' {\em MNRAS} {\bf 95} (Jan., 1935) 207--225.

\bibitem{1990RPPh...53..837K}
D.~{Koester} and G.~{Chanmugam}, ``{REVIEW: Physics of white dwarf stars},''
  {\em Reports on Progress in Physics} {\bf 53} (July, 1990) 837--915.

\bibitem{2015A&A...575A..60M}
G.~{Meynet}, V.~{Chomienne}, S.~{Ekstr{\"o}m}, C.~{Georgy}, A.~{Granada},
  J.~{Groh}, A.~{Maeder}, P.~{Eggenberger}, E.~{Levesque}, and P.~{Massey},
  ``{Impact of mass-loss on the evolution and pre-supernova properties of red
  supergiants},'' {\em Astronomy and Astrophysics} {\bf 575} (Mar., 2015) A60,
  \href{http://xxx.lanl.gov/abs/1410.8721}{{\tt 1410.8721}}.

\bibitem{Sasaki:2018dmp}
M.~Sasaki, T.~Suyama, T.~Tanaka, and S.~Yokoyama, ``{Primordial black holes:
  perspectives in gravitational wave astronomy},'' {\em Class. Quant. Grav.}
  {\bf 35} (2018), no.~6 063001, \href{http://xxx.lanl.gov/abs/1801.05235}{{\tt
  1801.05235}}.

\bibitem{Holberg:2001in}
J.~B. Holberg, T.~D. Oswalt, and E.~M. Sion, ``{A determination of the local
  density of white dwarf stars},'' {\em Astrophys. J.} {\bf 571} (2002) 512,
  \href{http://xxx.lanl.gov/abs/astro-ph/0102120}{{\tt astro-ph/0102120}}.

\bibitem{2017MNRAS.470.1360B}
J.~{Bovy}, ``{Stellar inventory of the solar neighbourhood using Gaia DR1},''
  {\em MNRAS} {\bf 470} (Sept., 2017) 1360--1387,
  \href{http://xxx.lanl.gov/abs/1704.05063}{{\tt 1704.05063}}.

\bibitem{Madau:2014bja}
P.~Madau and M.~Dickinson, ``{Cosmic Star Formation History},'' {\em Ann. Rev.
  Astron. Astrophys.} {\bf 52} (2014) 415--486,
  \href{http://xxx.lanl.gov/abs/1403.0007}{{\tt 1403.0007}}.

\bibitem{preparation}
G.~Cusin, I.~Dvorkin, C.~Pitrou, and J.-P. Uzan {\em in preparation} (2018).

\bibitem{Zumalacarregui:2017qqd}
M.~Zumalacarregui and U.~Seljak, ``{No LIGO MACHO: Primordial Black Holes, Dark
  Matter and Gravitational Lensing of Type Ia Supernovae},''
  \href{http://xxx.lanl.gov/abs/1712.02240}{{\tt 1712.02240}}.

\bibitem{Garcia-Bellido:2017imq}
J.~Garcia-Bellido, S.~Clesse, and P.~Fleury, ``{Primordial black holes survive
  SN lensing constraints},'' {\em Phys. Dark Univ.} {\bf 20} (2018) 95--100,
  \href{http://xxx.lanl.gov/abs/1712.06574}{{\tt 1712.06574}}.

\bibitem{Tisserand:2006zx}
{\bf EROS-2} Collaboration, P.~Tisserand {\em et.~al.}, ``{Limits on the Macho
  Content of the Galactic Halo from the EROS-2 Survey of the Magellanic
  Clouds},'' {\em Astron. Astrophys.} {\bf 469} (2007) 387--404,
  \href{http://xxx.lanl.gov/abs/astro-ph/0607207}{{\tt astro-ph/0607207}}.

\bibitem{Ricotti:2007au}
M.~Ricotti, J.~P. Ostriker, and K.~J. Mack, ``{Effect of Primordial Black Holes
  on the Cosmic Microwave Background and Cosmological Parameter Estimates},''
  {\em Astrophys. J.} {\bf 680} (2008) 829,
  \href{http://xxx.lanl.gov/abs/0709.0524}{{\tt 0709.0524}}.

\bibitem{Hawkins:2011qz}
M.~R.~S. Hawkins, ``{The case for primordial black holes as dark matter},''
  {\em Mon. Not. Roy. Astron. Soc.} {\bf 415} (2011) 2744,
  \href{http://xxx.lanl.gov/abs/1106.3875}{{\tt 1106.3875}}.

\bibitem{Green:2017qoa}
A.~M. Green, ``{Astrophysical uncertainties on stellar microlensing constraints
  on multi-Solar mass primordial black hole dark matter},'' {\em Phys. Rev.}
  {\bf D96} (2017), no.~4 043020,
  \href{http://xxx.lanl.gov/abs/1705.10818}{{\tt 1705.10818}}.

\bibitem{Garcia-Bellido:2017xvr}
J.~García-Bellido and S.~Clesse, ``{Constraints from microlensing experiments
  on clustered primordial black holes},'' {\em Phys. Dark Univ.} {\bf 19}
  (2018) 144--148, \href{http://xxx.lanl.gov/abs/1710.04694}{{\tt 1710.04694}}.

\bibitem{Poulin:2017bwe}
V.~Poulin, P.~D. Serpico, F.~Calore, S.~Clesse, and K.~Kohri, ``{CMB bounds on
  disk-accreting massive primordial black holes},'' {\em Phys. Rev.} {\bf D96}
  (2017), no.~8 083524, \href{http://xxx.lanl.gov/abs/1707.04206}{{\tt
  1707.04206}}.

\bibitem{Ali-Haimoud:2016mbv}
Y.~Ali-Haïmoud and M.~Kamionkowski, ``{Cosmic microwave background limits on
  accreting primordial black holes},'' {\em Phys. Rev.} {\bf D95} (2017), no.~4
  043534, \href{http://xxx.lanl.gov/abs/1612.05644}{{\tt 1612.05644}}.

\bibitem{Renzini:2018vkx}
A.~I. Renzini and C.~R. Contaldi, ``{Mapping Incoherent Gravitational Wave
  Backgrounds},'' \href{http://xxx.lanl.gov/abs/1806.11360}{{\tt 1806.11360}}.

\bibitem{Contaldi:2016koz}
C.~R. Contaldi, ``{Anisotropies of Gravitational Wave Backgrounds: A Line Of
  Sight Approach},'' {\em Phys. Lett.} {\bf B771} (2017) 9--12,
  \href{http://xxx.lanl.gov/abs/1609.08168}{{\tt 1609.08168}}.

\bibitem{Cusin:2016kqx}
G.~Cusin, C.~Pitrou, and J.-P. Uzan, ``{Are we living near the center of a
  local void?},'' {\em JCAP} {\bf 1703} (2017), no.~03 038,
  \href{http://xxx.lanl.gov/abs/1609.02061}{{\tt 1609.02061}}.

\bibitem{deSalas:2016ztq}
P.~F. de~Salas and S.~Pastor, ``{Relic neutrino decoupling with flavour
  oscillations revisited},'' {\em JCAP} {\bf 1607} (2016), no.~07 051,
  \href{http://xxx.lanl.gov/abs/1606.06986}{{\tt 1606.06986}}.

\bibitem{Ade:2015xua}
{\bf Planck} Collaboration, P.~A.~R. Ade {\em et.~al.}, ``{Planck 2015 results.
  XIII. Cosmological parameters},'' {\em Astron. Astrophys.} {\bf 594} (2016)
  A13, \href{http://xxx.lanl.gov/abs/1502.01589}{{\tt 1502.01589}}.

\bibitem{Marin:2013bbb}
{\bf WiggleZ} Collaboration, F.~A. Marin {\em et.~al.}, ``{The WiggleZ Dark
  Energy Survey: constraining galaxy bias and cosmic growth with 3-point
  correlation functions},'' {\em Mon. Not. Roy. Astron. Soc.} {\bf 432} (2013)
  2654, \href{http://xxx.lanl.gov/abs/1303.6644}{{\tt 1303.6644}}.

\bibitem{Rassat:2008ja}
A.~Rassat, A.~Amara, L.~Amendola, F.~J. Castander, T.~Kitching, M.~Kunz,
  A.~Refregier, Y.~Wang, and J.~Weller, ``{Deconstructing Baryon Acoustic
  Oscillations: A Comparison of Methods},''
  \href{http://xxx.lanl.gov/abs/0810.0003}{{\tt 0810.0003}}.

\end{thebibliography}\endgroup

\end{document}